\documentclass[amsfonts]{revtex4}
\usepackage{graphicx}
\begin{document}

\title{Scalar-field perturbations from a particle orbiting a
black hole using numerical evolution in 2+1 dimensions}

\author{Leor Barack and Darren A Golbourn}
\affiliation
{School of Mathematics, University of Southampton, Southampton,
SO17 1BJ, United Kingdom}

\date{\today}

\begin{abstract}
We present a new technique for time-domain numerical evolution of the scalar field generated by a pointlike scalar charge orbiting a black hole.  Time-domain evolution offers an efficient way for calculating black hole perturbations, especially as input for computations of the local self force acting on orbiting particles. In Kerr geometry, the field equations are not fully separable in the time domain, and one has to tackle them in 2+1 dimensions (two spatial dimensions and time; the azimuthal dependence is still separable). A technical difficulty arises when the source of the field is a pointlike particle, as the 2+1-dimensional perturbation is then singular: Each of the azimuthal modes diverges logarithmically at the particle.  To deal with this problem we split the numerical domain into two regions:  Inside a thin worldtube surrounding the particle's worldline we solve for a regularized variable, obtained from the full field by subtracting out a suitable ``puncture'' function, given analytically.  Outside this worldtube we solve for the full, original field. The value of the evolution variable is adjusted across the boundary of the worldtube.  In this work we demonstrate the applicability of this method in the example of circular orbits around a Schwarzschild black hole (refraining from exploiting the spherical symmetry of the background, and working in 2+1 dimensions).
\end{abstract}

\maketitle

\section{Introduction}
The motivation for this work stems from the problem of calculating the gravitational self-force (SF) acting on mass particles in orbit around black holes. This problem has drawn much attention recently, in relation with the effort to model the inspiral of compact objects into massive black holes in galactic nuclei---one of the prime targets for LISA, the planned space-based gravitational-wave detector. The scientific merit from detecting such inspirals is potentially high \cite{Barack:2003fp}, but full exploitation of the gravitational-wave signal will require precise knowledge of the theoretical phase evolution of the waves, as predicted by general relativity, over a few years of inspiral.  This, in turn, will require knowledge of the orbital evolution over a similar timescale, which, for sources of interest for LISA, would mean that the effects of the gravitational self force will have to be accounted for in the model. Thanks to the small mass ratio characteristic of the relevant binary systems ($\mu/M=10^{-4}$--$10^{-7}$, where $\mu$ is the mass of the compact object and $M$ is the mass of the massive black hole), the problem can be studied within the realm of perturbation theory, i.e., by considering the small perturbation caused by the SF to the orbit of a test particle moving in the fixed geometry of the central black hole.

There is now a well established theoretical framework for SF calculations in curved spacetimes \cite{DeWitt:1960fc,Mino:1996nk,Quinn:1996am,Quinn:2000wa,Detweiler:2002mi}, along with a practical calculation scheme for particle orbits around Kerr black holes \cite{Barack:1999wf,BMNOS2002,Barack:2002mh}. This method, dubbed ``mode-sum scheme'', requires as input the local metric perturbation (or its multipole modes) near the particle, in a particular gauge---the Lorenz gauge. In the special case of a Schwarzschild spacetime (i.e., a non-rotating central hole), the Lorenz-gauge metric perturbation can be obtained by solving the linearized Einstein equations directly, for each tensor-harmonic mode of the perturbation (using, e.g., numerical evolution in the time domain, as in \cite{BL2005}). This method was applied recently for circular orbits in Schwarzschild, allowing a first calculation of the local gravitational SF for an orbiting particle \cite{BS2007}. [Earlier calculations considered radial infall trajectories \cite{Barack:2002ku} and static (supported) particles \cite{KFW2006}, neither scenarios likely to be of relevance to LISA.]

The work presented here is a first step towards tackling the problem in the Kerr spacetime.  To obtain the Lorenz-gauge perturbation in Kerr, one may follow one of two possible avenues of approach. In the first approach, one first applies the Teukolsky formalism to solve for the perturbation in the Weyl scalars (fully decoupled into Fourier-harmonic modes if one opts to work in the frequency domain), and then reconstructs the corresponding metric perturbation in the Lorenz gauge. A procedure for Lorenz-gauge metric reconstruction is yet to be devised\footnote{A procedure to obtain the SF directly from the Weyl scalars through direct regularization of the latter has been suggested recently, but has not yet been implemented for orbiting particles. \cite{KFW2006}}. The alternative approach, which we pursue here, is to solve directly for the metric perturbation, as in Refs.\ \cite{BL2005,BS2007}, using time-domain numerical evolution of the Lorenz-gauge perturbation equations. This method offers a few important advantages: First, the problem of reconstructing the metric perturbation is avoided.
Second, the behavior of the Lorenz-gauge perturbation near the particle, unlike that of the Weyl scalars, reflects the physical, isotropic form of the particle singularity, and is therefore more tractable. Third, the time-domain treatment best exploits the hyperbolic nature of the Lorenz-gauge perturbation equations. Finally, the time-domain treatment makes it easier to tackle particle orbits of arbitrary eccentricity.

The main challenge in applying the above approach relates to the fact that the perturbation equations in Kerr spacetime are not fully separable in the time domain. One can at most separate out the azimuthal dependence, then consider the evolution problem for each of the resulting `$m$-modes', each of which being a field of 2+1 dimensions (2+1-D), depending on time and on 2 spatial coordinates. The non-separability of the field equations,
on its own, does not pose a serious problem: Evolution codes for vacuum perturbations in 2+1-D have been developed and used successfully since the mid-1990s \cite{KW1996,KW1997, LAR2003,KG2003,PAL2004,SKH2007}. The difficulty, rather, arises from the inclusion of a point particle as a source for the perturbation. Each $m$-mode of the resulting perturbation then diverges (logarithmically) at the particle, and accommodating this physical singularity on the discrete numerical grid becomes a major concern and the main challenge.

Our goal here is to develop a scheme for handling the particle singularity within a 2+1-D evolution code. The idea is simple, and can be described as follows. As in Ref.\ \cite{BS2007} (and unlike in \cite{SKH2007}), we model the orbiting particle with a spatial delta-function distribution. The asymptotic form of the local perturbation field near the particle is then known analytically (e.g., \cite{BO2003}). We then construct a function (given analytically) which (i) has
the same local asymptotic form and (ii) is easily decomposed into $m$-modes. The difference between the full perturbation and this ``singular'' function defines a new, ``regularized'' field. The singular function is so designed that each of the $m$-modes of the regularized field is continuous at the location of the particle. We then use the ($m$-modes of the) regularized field as our numerical evolution variables. The full solution is simply the sum of the
numerically-calculated regular field, and the analytically-given singular function. (Our ``regular'' function is, by construction, continuous, but not necessarily smooth; The regular function to be constructed in this work will have discontinuous derivatives at the particle's location. It should be stressed, in this regard, that our ``regular'' and ``singular'' variables do {\it not} necessarily correspond to Detweiler and Whiting's `R' and `S' fields \cite{Detweiler:2002mi}, the latter so defined that the `R' field is a homogeneous, smooth solution of the perturbation equations.)

To make it easier to control the global properties of our numerical evolution variable (especially its behavior at infinity and along the horizon), our evolution code will apply the above procedure only at the vicinity of the particle; far away from the particle it will utilize as an integration variable the full, original homogeneous field. To make this work in practice, we will introduce a reference ``worldtube'' around the worldline (in the 2+1-D space of the $m$-mode fields), whose ``width'' will be taken to be of order the background's radius of curvature, but will otherwise remain a control parameter in our numerical code. At each time step of the numerical evolution, the code will solve for the regular field inside the wordtube and for the original full field outside it, simply adjusting the value of the numerical variable across the boundary of the worldtube (using the known difference between the full and regular fields, being just the
value of the singular function). In validating the numerical code, it will be important to monitor the amount by which the numerical solutions depend on the worldtube dimensions, and verify that this dependence becomes negligible with increasing grid resolution.

The idea of representing a singular part of the solution analytically and solving numerically for the remaining regular part is reminiscent of the ``puncture'' method, often used in Numerical Relativity in representing initial data for spacetimes containing black holes \cite{BB1997}.  We shall call our singular variable a ``puncture function'', and refer to our scheme as the ``puncture method'', but we remind that here the idea of a puncture is applied in a different physical context.

In this manuscript we demonstrate the applicability of the above method using a simple scalar-field toy model. To simplify the analysis still, we will consider circular orbits around a Schwarzschild, rather than Kerr, black hole. However, we will refrain from exploiting the spherical symmetry of the background geometry, pretending that the field equations cannot be separated into spherical-harmonic modes, and working in 2+1-D.  The code we develop here should be expandable in a rather direct way to the Kerr spacetime and to eccentric/inclined orbits. We envisage applying a similar procedure for gravitational perturbations, but this would require much more development, including the formulation of the Lorenz-gauge perturbation equations is a format suitable for numerical evolution in 2+1-D.

The structure of this paper is as follows. In Sec.\ \ref{Sec:II} we decompose the scalar field in Schwarzschild spacetime into $m$-modes, and analyze the behavior of the individual modes near the particle, showing the logarithmic divergence.  In Sec.\ \ref{Sec:III} we formulate our puncture scheme, select a particular puncture function, and analyze the asymptotic behavior of the regular field near the particle. In Sec.\ \ref{Sec:IV} we describe our 2+1-D numerical evolution code as applied for vacuum perturbations. We test it for numerical convergence, and
check that, for initial perturbations with compact support, the late-time decay rate of individual multipole modes agree with that predicted by theory. We also test the 2+1-D vacuum solutions against solutions obtained with a 1+1-D code. In Sec.\ \ref{Sec:V} we use the puncture scheme to incorporate a source term in our code, representing a scalar-charge particle moving in a circular geodesic orbit. We detail the numerical procedure in this case.  Sec.\ \ref{Sec:VI} gives some results for the particle case, and presents a list of validation tests for the code. These include (i) test of point-wise numerical convergence, (ii) test of independence on the worldtube dimensions, and (iii) comparison with solutions obtained using a 1+1-D code. In Sec.\ \ref{Sec:VII} we summarize this work, and discuss the application of our method for SF calculations.

In passing, we briefly mention some related literature. Over the past decade, several authors have considered the evolution of black hole perturbations in 2+1-D, with or without a particle source. Krivan \emph{et al.\ }\cite{KW1996} wrote a 2+1-D code to analyze the late-time power-law decay of homogeneous scalar field perturbations in Kerr spacetimes. Krivan \emph{et al.\ }\cite{KW1997} later examined also the late-time dynamics of the Weyl scalars associated with vacuum gravitational perturbations, by solving Teukolsky's master equation in 2+1-D. More recently, Pazos-Avalos and Lousto\cite{PAL2004} presented an improved, fourth-order-convergent code in 2+1-D, for the evolution of vacuum perturbations of the Teukolsky equation. Particle orbits in Kerr have been tackled with a 2+1-D code by Lopez-Aleman \emph{et al.\ }\cite{LAR2003}, Khanna \cite{KG2003}, and, more recently, Burko and Khanna \cite{BK2007} and Sundararajan \emph{et al.\ }\cite{SKH2007}.  In these works (reporting on a series of improvements to the same 2+1-D code for evolution of the Teukolsky equation), the particle is represented by a smeared distribution of matter. The most recent of this works has achieved a reasonable accuracy in the far-field solutions, but the method is likely inadequate for accurate determination of the local field near the particle, which is essential for SF calculations.  Sopuerta and Laguna \cite{SSLX2006} suggested the use of finite-element methods for an effective treatment of the particle. This idea (so far implemented for orbits in
Schwarzschild \cite{SL2006}) shows much promise, but requires more development.  Finally, Bishop \emph{et al.\ }\cite{BGHLW2003} have tackled the extreme-mass-ratio inspiral problem within the framework of full numerical relativity, i.e., by solving the full non-linear Einstein equations. This approach, too, requires more development.

Throughout this paper we use metric signature $=\mathrm{diag}(-,+,+,+)$, and work in geometrised units, with $G=c=1$.

\section{Decomposition of the scalar field in 2+1-D} \label{Sec:II}
\subsection{Physical setup: scalar particle in Schwarzschild}
\label{subsec:setup}
We consider a pointlike test particle endowed with scalar charge $q$, moving in a circular orbit around a Schwarzschild black hole of mass $M$. In this work we ignore the SF, and assume the particle moves on a circular geodesic of the Schwarzschild background. Let $x^{\alpha}_{\rm p}(\tau)$ denote the worldline of the particle (parametrized by proper time $\tau$), and introduce the tangent four-velocity $u^{\alpha}=dx^{\alpha}_{\rm p}/d\tau$. Without loss of generality we work in a standard Schwarzschild coordinate system $(t,r,\theta,\varphi)$ in which the orbit is confined to the equatorial plane, $\theta_p=\pi/2$. We then have
\begin{equation}\label{dalem}
x^{\alpha}_{\rm p}=[t_{\rm p}(\tau),r_0={\rm const},\pi/2,\omega t_{\rm p}(\tau)],
\quad\quad
u^{\alpha}_{\rm p}={\cal E}/f_0[1,0,0,\omega],
\end{equation}
where $r_0$ is the orbital `radius',
\begin{equation}\label{omega}
\omega \equiv d\varphi_{\rm p}/dt_{\rm p}=(M/r_0^3)^{1/2}
\end{equation}
is the angular frequency (with respect to time $t$), and
\begin{equation}\label{E}
{\cal E}\equiv -u_{{\rm p}t}=f_0(1-3M/r_0)^{-1/2}
\end{equation}
is the specific energy parameter, with $f_0\equiv 1-2M/r_0$.

We take the scalar field $\Phi$ of the particle to be minimally-coupled and massless.  It then satisfies
\begin{equation}\label{dalem2}
\Box\Phi \equiv \frac{1}{\sqrt{-g}}\left(g^{\alpha\beta} \Phi_{,\beta}\sqrt{-g}\right)_{,\alpha}
= S,
\end{equation}
where $g^{\alpha\beta}$ represents the background (Schwarzschild) metric, $g$ is the background metric determinant, and the source term is given by
\begin{eqnarray} \label{source}
S &\equiv& -4\pi q\int^{\infty}_{-\infty} \frac{\delta^4\left[x - x_{\rm p}(\tau)\right]}
{\sqrt{-g}}\mathrm{d}\tau \nonumber\\
&=& \frac{-4\pi q}{r_0^2}\frac{f_0}{\cal E}\delta(r-r_0)\delta(\theta-\frac{\pi}{2})
\delta(\varphi-\omega t_{\rm p}) .
\end{eqnarray}

\subsection{$m$-mode decomposition}
To reduce the problem to 2+1-D we decompose $\Phi$ into azimuthal modes, in the form

\begin{equation}\label{ansatz21}
\Phi = \sum_{m=-\infty}^{\infty} e^{im\varphi}\Phi^{m}(t,r,\theta) .
\end{equation}
The individual $m$-modes are obtained through
\begin{equation}\label{Phim}
\Phi^{m}=\frac{1}{2\pi}\int_{-\pi}^{\pi}\Phi e^{-im\varphi}d\varphi.
\end{equation}
Note that, for future convenience, we take the principal values of the coordinate $\varphi$ to lie in the range $-\pi<\varphi\leq\pi$. The scalar field equation (\ref{dalem}) separates as
\begin{eqnarray}\label{w22lhs}
\Box^m\Phi^m \equiv g^{tt}\Phi^m_{,tt} + g^{rr}\Phi^m_{,rr}
+\left(f_{,r} + 2r^{-1}g^{rr}\right)\Phi^m_{,r}
+g^{\theta\theta}\left(\Phi^m_{,\theta\theta} + \cot\theta\,\Phi^m_{,\theta}\right)
-m^2g^{\phi\phi}\Phi^m =S^m,
\end{eqnarray}
where the $m$-mode source reads
\begin{equation}\label{w22rhs}
S^m = \frac{-4\pi q}{r_0^2}(1-3M/r_0)^{1/2}\delta(r-r_0)
\delta(\theta-\frac{\pi}{2})e^{-im\omega t_{\rm p}}.
\end{equation}
We note the relation $(\Phi^m)^*=\Phi^{-m}$ (where an asterix denotes complex conjugation), which allows us to fold the $m<0$ part of the sum in Eq.\ (\ref{ansatz21}) over onto $m>0$:
\begin{equation}
\Phi=\Phi^{m=0}+2\sum_{m=1}^{\infty}{\rm Re}\left(e^{im\varphi}\Phi^m\right).
\end{equation}

To cast Eq.\ (\ref{w22lhs}) in a form more suitable for numerical integration, we introduce the new variable
\begin{equation}\label{ret}
\Psi^m=r\Phi^m.
\end{equation}
In terms of $\Psi^m$, the field equation takes the form
\begin{eqnarray}\label{psis}
\Box_{\Psi}^m\Psi^m\equiv \Psi^m_{,uv}-\frac{f}{4r^2}\left[
\Psi^m_{,\theta\theta} + \cot\theta\,\Psi^m_{,\theta}
-\left(2M/r+m^2\csc^2\theta\right)\Psi^m\right]=-(fr/4)S^m,
\end{eqnarray}
where $f\equiv 1-2M/r$, and $u$ and $v$ are the standard Eddington-Finkelstein null coordinates (`retarded' and `advanced'-time coordinates, respectively),
given by
\begin{equation}\label{uv}
u = t - r_*, \quad\quad v = t+ r_*,
\end{equation}
with
\begin{equation}
r_* = r + 2M\ln\left(\frac{r-2M}{2M}\right).
\end{equation}

\subsection{Behavior of $\Phi^m$ near the particle}
We now show that each of the modes $\Phi^m$ diverges as $x\to x_{\rm p}$, and that this divergence is logarithmic (in the proper distance).

The singular behavior of the full scalar field near the particle is known to be described, at leading order, by \cite{BO2002,MNS2003}
\begin{equation}\label{phis1}
\Phi(x) \simeq \frac{q}{\epsilon}.
\end{equation}
Here $x$ represents a point near the worldline, and $\epsilon$ is the spatial geodesic distance from $x$ to the worldline, i.e., the length of the small geodesic section connecting $x$ to the worldline and normal to it. If $x_{\rm p}$ is a worldline point near $x$ (not necessarily the intersection of the above normal geodesic with the worldline), and $\delta x^\alpha \equiv x^\alpha - x_{\rm p}^\alpha$, then, at leading order in the coordinate distance, $\epsilon$ is given by
\begin{equation}\label{epsilon0}
\epsilon^2 \simeq P_{\alpha\beta}\delta x^\alpha \delta x^\beta,
\end{equation}
where $P_{\alpha\beta}$ is a spatial projection operator reading
\begin{equation}\label{P}
P_{\alpha\beta}=g_{\alpha\beta}(x_{\rm p}) + u_\alpha(x_{\rm p}) u_\beta(x_p).
\end{equation}

Consider now a particular point $x_{\rm p}$ on the worldline, and let $\Sigma$ be the spatial hypersurface $t=t_{\rm p}$, containing $x_{\rm p}$. In the following we consider points $x$ on $\Sigma$, and ask how $\Phi^m(x)$ behaves as $x\to x_{\rm p}$. Specializing to circular equatorial orbits in Schwarzschild, we take, for simplicity (but with no loss of generality), $\varphi_{\rm p}=0$, and introduce local polar coordinates $\rho,\phi$ in the $r$--$\theta$ plane:
\begin{eqnarray}\label{polars}
\delta r&=&r-r_0=P_{rr}^{-1/2}\rho\cos\phi, \nonumber\\
\delta \theta&=&\theta-\theta_0=P_{\theta\theta}^{-1/2}\rho\sin\phi.
\end{eqnarray}
Then, at leading order,
$\epsilon_{\Sigma}=\left(\rho^2+P_{\varphi\varphi}\varphi^2\right)^{1/2}$,
where the subscript $\Sigma$ reminds us that $\epsilon$ is evaluated on $\Sigma$, i.e, at $t=t_{\rm p}$. Substituting $\Phi\simeq q/\epsilon_{\Sigma}$ in Eq.\ (\ref{Phim}), we obtain
\begin{equation}\label{PhimSing}
\Phi^{m}(\rho)\simeq\frac{1}{2\pi}\int_{-\pi}^{\pi}
\frac{e^{-im\varphi}}{\left(\rho^2+P_{\varphi\varphi}\varphi^2\right)^{1/2}}\,d\varphi
\quad \text{(for $x\to x_{\rm p})$},
\end{equation}
which describes the asymptotic behavior of $\Phi^m$ as one approaches the worldline along a $t$=const trajectory. Note that the particle limit corresponds to $\rho\to 0$.

To evaluate the above integral, we write it in the form
\begin{equation}\label{split}
\int_{-\pi}^{\pi}
\frac{e^{-im\varphi}}{\left(\rho^2+P_{\varphi\varphi}\varphi^2\right)^{1/2}}\,d\varphi
=
\int_{-\pi}^{\pi}
\frac{e^{-im\varphi}-1}{\left(\rho^2+P_{\varphi\varphi}\varphi^2\right)^{1/2}}\,d\varphi
+
\int_{-\pi}^{\pi}
\frac{1}{\left(\rho^2+P_{\varphi\varphi}\varphi^2\right)^{1/2}}\,d\varphi.
\end{equation}
Using $\left|e^{-im\varphi}-1 \right|=\sqrt{2(1-\cos m\varphi)}\leq m|\varphi|$ (valid for $|\varphi|\leq\pi$), the magnitude of the first integral on the right-hand side can be bounded, for any fixed value of $\rho$, as
\begin{equation}\label{bound2}
\left|\int_{-\pi}^{\pi}
\frac{e^{-im\varphi}-1}{\left(\rho^2+P_{\varphi\varphi}\varphi^2\right)^{1/2}}\,d\varphi
\right|
\leq
\int_{-\pi}^{\pi}
\frac{m|\varphi|}{\left(\rho^2+P_{\varphi\varphi}\varphi^2\right)^{1/2}}\,d\varphi
\leq
\int_{-\pi}^{\pi}
\frac{m}{P_{\varphi\varphi}^{1/2}}\,d\varphi
=
\frac{2\pi m}{P_{\varphi\varphi}^{1/2}}
\end{equation}
(recalling $P_{\varphi\varphi}>0$). Hence, this contribution is bounded at the limit $\rho\to 0$.
Consider next the contribution from the second integral on the right-hand side of Eq.\ (\ref{split}). It gives
\begin{eqnarray}\label{bound3}
\int_{-\pi}^{\pi}
\frac{d\varphi}{\left(\rho^2+P_{\varphi\varphi}\varphi^2\right)^{1/2}}&=&
\frac{\pi}{\rho_0}\ln\left[
\frac{(\rho_0^2+\rho^2)^{1/2}+\rho_0}{(\rho_0^2+\rho^2)^{1/2}-\rho_0} \right] \nonumber\\
&=&
-\frac{2\pi}{\rho_0}\ln\left(\frac{\rho}{2\rho_0}\right)+
O(\rho^2),
\end{eqnarray}
where $\rho_0\equiv \pi P_{\varphi\varphi}^{1/2}$ (depending on $r_0$ only), and where in the last step we expanded about $\rho\to 0$.

Collecting the results (\ref{bound2}) and (\ref{bound3}), we conclude that, at leading order,
\begin{equation}\label{Asymp}
\Phi^m(\rho\to 0)=
-\rho_0^{-1}\ln\left(\frac{\rho}{2\rho_0}\right),
\end{equation}
i.e., each of the $m$-modes of the scalar field diverges {\em logarithmically} with $\rho$, approaching the particle.  (Note that $\rho$ is the proper distance along geodesics in $\Sigma$ emanating `radially' from the particle.) Interestingly, the form of the leading-order divergence does not depend on the mode number $m$.

Although we have restricted the above discussion to circular orbits in Schwarzschild, it is straightforward to repeat the analysis with an arbitrary point $x_{\rm p}$ along an arbitrary geodesic orbit in Kerr. The main conclusion holds in general:
$\Phi^m\propto\ln\rho$ as $\rho\to 0$, for any $m$.

\section{Puncture scheme} \label{Sec:III}

The divergence of $\Phi^m$ along the worldline is a serious concern when considering the numerical integration of the scalar field in 2+1-D. To deal with this difficulty, we introduce the following scheme.

The scheme involves the introduction of a scalar field $\Phi_{\rm P}$ (`P' for {\it puncture}), given analytically, whose singular structure is similar to that of $\Phi$. More precisely, we choose the field $\Phi_{\rm P}$ such that each azimuthal $m$-mode of the difference
\begin{equation}
\Phi_{\rm R}\equiv \Phi-\Phi_{\rm P}
\end{equation}
is {\em bounded} and {\em continuous} at the worldline.  We then use the $m$-modes of $\Phi_{\rm R}$ as new variables for the numerical integration in a region near the worldline.
Specifically, we introduce a worldtube surrounding the worldline, the dimensions of which are kept controllable numerical parameters. Let $\cal T$ denote the interior of this worldtube, and $\partial {\cal T}$ its boundary. Let also $\Phi^m_{\rm R}$ and $\Phi^m_{\rm S}$ denote the $m$-modes of $\Phi_{\rm R}$ and $\Phi_{\rm P}$, respectively:
\begin{equation}\label{PhiPRm}
\Phi^{m}_{\rm R}=\frac{1}{{2\pi}}\int_{-\pi}^{\pi}\Phi_{\rm R} e^{-im\varphi}d\varphi,
\quad\quad
\Phi^{m}_{\rm P}=\frac{1}{{2\pi}}\int_{-\pi}^{\pi}\Phi_{\rm P} e^{-im\varphi}d\varphi.
\end{equation}
The numerical scheme then utilizes the ``regularized'' variables $\Phi^m_{\rm R}$ for the part of the evolution which takes place inside $\cal T$, while outside $\cal T$ it evolves the original fields $\Phi^m$. The value of the evolution variable is adjusted across $\partial {\cal T}$ using
$\Phi^m_{\rm R}= \Phi^m-\Phi^m_{\rm P}$.
Thus, within our puncture scheme, the field equations to be evolved are
\begin{equation}\label{scheme}
\left\{ \begin{array}{ll}
\Box^m \Phi^m_{\rm R}=S^m-\Box^m \Phi^m_{\rm P}\equiv S^m_{\rm R}
                & {\rm in\ } \cal T, \\
\Box^m \Phi^m=0
                & {\rm outside\ } \cal T, \\
{\rm with \ }\Phi^m_{\rm R}= \Phi^m-\Phi^m_{\rm P}
                & {\rm on\ } \partial \cal T, \\
\end{array}
\right.
\end{equation}
where $S^m_{\rm R}$ and $\Phi^m_{\rm P}$ are given analytically.  Of course, once the continuous fields $\Phi^m_{\rm R}$ are solved for, the full scalar-field modes can be simply constructed through $\Phi^m=\Phi^m_{\rm R}+\Phi^m_{\rm P}$.  Note that, depending on the form of the puncture function $\Phi_{\rm P}$, the source $S^m_{\rm R}$ may have support anywhere inside $\cal T$ (not necessarily confined to the worldline).

\subsection{Choice of the puncture function $\Phi_{\rm P}$}
We wish to construct a function $\Phi_{\rm P}$ which (i) reproduces the singular behavior of the full field $\Phi$ at $x\to x_{\rm p}$; (ii) is sufficiently regular away from the particle; and (iii) is easily decomposable, in analytic form, into $m$-modes.

Consider the puncture function
\begin{equation}\label{puncturefn}
\Phi_{\rm P}(x;x_{\rm p}) =
\frac{q}{\epsilon_{\rm P}},
\end{equation}
with
\begin{equation}\label{epsilonP}
\epsilon_{\rm P}=\sqrt{\rho^2 + 2P_{\varphi\varphi}(1-\cos\delta\varphi)}\ .
\end{equation}
Here, as in Sec.\ \ref{Sec:II}, $\delta x^{\alpha}\equiv x^{\alpha}-x_{\rm p}^{\alpha}$, $P_{\alpha\beta}$ are tensorial coefficients as defined in Eq.\ (\ref{P}), and $\rho$ [same as in Eq.\ (\ref{polars})] is given explicitly by
\begin{equation}\label{rho}
\rho^2=P_{rr}\delta r^2+P_{\theta\theta}\delta\theta^2.
\end{equation}
Note that here we do not regard the coordinate differences $\delta x$ as necessarily small.  The definitions in Eqs.\ (\ref{puncturefn})--(\ref{rho}) are taken as {\em exact}, for any value of $\delta x$ within the worldtube $\cal T$.

Since $2(1-\cos\delta\varphi)=\delta\varphi^2+O(\delta\varphi^4)$, the function $\epsilon_{\rm P}$ coincides, at leading order in $\delta x$, with the function $\epsilon$ [see Eq.\ (\ref{epsilon0})], evaluated on the hypersurface $t=t_{\rm p}$.  Therefore, at leading order in $\delta x$, the puncture function $\Phi_{\rm P}$ coincides with $\Phi(t=t_{\rm p})$. As desired, $\Phi_{\rm P}$ is singular only at the location of the particle ($\rho=\delta\varphi=0$), and is regular ($C^{\infty}$) anywhere else. Finally, as we show below, our function $\Phi_{\rm P}$ is easily decomposed, in explicit form, into $m$-modes.\footnote{One may consider an alternative puncture
function, obtained by replacing $2(1-\cos\delta\varphi)\to \sin^2\delta\varphi$ in Eq.\ (\ref{epsilonP}). This has the advantage that the odd-$m$ modes of $\Phi_{\rm P}$ and $S_{\rm R}$ can be written in terms of elementary functions. The disadvantage is that, to avoid the singularity of this alternative $\Phi_{\rm P}$ at $\rho=0$, $\delta\varphi=\pi$, one has to introduce a cut-off on $\Phi_{\rm P}$ at some $|\delta\varphi|<\pi$, which then generates distributional contributions to the source modes $S_{\rm R}^m$, complicating their form considerably.}

\subsection{Continuity of the modes $\Phi_{\rm R}^m$}
We now show that, with the puncture function selected above, the modes $\Phi^m_{\rm R}\equiv \Phi^m-\Phi^m_{\rm P}$ are finite and continuous for all $r$ and $\theta$.  Since both $\Phi$ and $\Phi_{\rm P}$ are regular ($C^\infty$) away from the particle, then so is $\Phi_{\rm R}$, and so are its modes $\Phi^m_{\rm R}$. We hence focus on the behavior of the modes $\Phi^m_{\rm P}$ at the location of the particle ($\delta r=\delta\theta=0$, or, equivalently, $\rho=0$), aiming to show that they are $C^{0}$ there.

For this discussion, we will need to consider higher-order terms in the asymptotic formula (\ref{phis1}). It was shown in Ref.\ \cite{MNS2003} (by considering the Hadamard expansion of the retarded Green's function for the scalar field) that, near the particle,
\begin{equation} \label{Phi}
\Phi(x)=\frac{q}{\epsilon}+f_1(x),
\end{equation}
where $f_1$ is a $C^0$ function (i.e., continuous but not necessarily differentiable). The spatial geodesic distance $\epsilon$ can be expanded in terms of the coordinate difference $\delta x^{\alpha}$ in the form
\begin{equation}\label{epsilon}
\epsilon^2=\epsilon_{0}^2
          +Q_{\alpha\beta\gamma}(x_{\rm p})\delta x^{\alpha}\delta x^{\beta}\delta x^{\gamma}
          +O(\delta x^4),
\end{equation}
where $\epsilon_0^2=P_{\alpha\beta}(x_{\rm p})\delta x^{\alpha}\delta x^{\beta}$ [the leading-order form, as in Eq.\ (\ref{epsilon0})], and $Q_{\alpha\beta\gamma}$ are certain coefficients depending only on $x_{\rm p}$ (they are given explicitly, e.g., in Ref.\ \cite{BO2002}). Substituting for $\epsilon$ from Eq.\ (\ref{epsilon}), Eq.\ (\ref{Phi}) becomes
\begin{equation}\label{2ndorder}
\Phi(x)= \frac{q}{\epsilon_0}
-\frac{1}{2}qQ_{\alpha\beta\gamma}\frac{\delta x^{\alpha}\delta x^{\beta}\delta x^{\gamma}}
{\epsilon_0^{3}}+f_2(x),
\end{equation}
where $f_2$ is $C^0$.

In what follows we fix $x_{\rm p}$, and consider the behavior of $\Phi$ (and $\Phi_{\rm P}$) on the hypersurface $t=t_{\rm p}$, denoted $\Sigma$ as before. Recalling the definition of
$\epsilon_{\rm P}$ in Eq.\ (\ref{epsilonP}), we have, on $\Sigma$,
\begin{equation}
\epsilon_{\rm P}^2=\epsilon_{0}^2 +O(\delta x^4)
\end{equation}
and hence (near the particle)
\begin{equation}
\Phi_{\rm P}=\frac{q}{\epsilon_0}+O(\delta x).
\end{equation}
Thus, on $\Sigma$,
\begin{equation} \label{PhiRasy}
\Phi_{\rm R}=\Phi-\Phi_{\rm P}=
-\frac{1}{2}qQ_{\alpha\beta\gamma}\frac{\delta x^{\alpha}\delta x^{\beta}\delta x^{\gamma}}
{\epsilon_0^{3}}+f_3(x),
\end{equation}
where $f_3$ is yet another $C^0$ function. In this expression $\delta x^{\alpha}=\delta r$, $\delta\theta$, or $\delta\varphi$, and
\begin{equation}
\epsilon_0=(\rho^2+P_{\varphi\varphi}\delta\varphi^2)^{1/2},
\end{equation}
where, recall, $\rho$ is given in Eq.\ (\ref{rho}). We now write $\Phi_{\rm R}e^{-im\delta\varphi}=\Phi_{\rm R}[1+O(\delta\varphi)]$ (for any $m$, at small $|\delta\varphi|$), and notice that, by virtue of Eq.\ (\ref{PhiRasy}), the contribution to $\Phi_{\rm R}e^{-im\delta\varphi}$ from the
$O(\delta\varphi)$ terms vanishes as $\epsilon_0\to 0$. Hence, this contribution is $C^0$, and we may write
\begin{equation}
\Phi_{\rm R}e^{-im\delta\varphi}=
-\frac{1}{2}qQ_{\alpha\beta\gamma}\frac{\delta x^{\alpha}\delta x^{\beta}\delta x^{\gamma}}
{\epsilon_0^{3}}+f_4^m(x),
\end{equation}
where $f_4^m(x)$ is a $C^0$ function (depending on $m$), and where the first term on the RHS is $m$-independent. For simplicity (but without loss of generality) we take $\varphi_{\rm p}=0$, giving $\delta\varphi=\varphi$. The $m$ modes of $\Phi_{\rm R}$ are then given by
\begin{equation} \label{reg1}
\Phi_{\rm R}^m=
-\frac{1}{2}qQ_{\alpha\beta\gamma}\int_{-\pi}^{\pi}
\frac{\delta x^{\alpha}\delta x^{\beta}\delta x^{\gamma}}
{\epsilon_0^{3}}d\varphi+f_5^m,
\end{equation}
where the integral $f_5^m(r,\theta)\equiv \int_{-\pi}^{\pi}f_4^m d\varphi$ is necessarily a $C^0$ function of $\delta r$ and $\delta\theta$ (since the integrand $f_4^m$ is a $C^0$ function of $\delta r$, $\delta\theta$ and $\varphi$).

It remains to show that the term $\propto Q_{\alpha\beta\gamma}$ in Eq.\ (\ref{reg1}) is $C^0$. (Note that the integrand in this term is not necessarily $C^0$.) To this end, we write $(-q/2)Q_{\alpha\beta\gamma}\delta x^{\alpha}\delta x^{\beta} \delta x^{\gamma}$ explicitly as a polynomial in $\varphi$, in the form $p_3+p_2\varphi+p_1\varphi^2+p_0\varphi^3$, were $p_n(\delta r,\delta\theta)$ are polynomials in $\delta r$ and $\delta\theta$, each of the form
$\sum_{k=0}^n a_k \delta r^k \delta\theta^{n-k}$ (with $a_k$ constant coefficients). The contributions from the terms $\propto p_2,p_0$ to the integral in Eq.\ (\ref{reg1}) vanish from symmetry. The contribution from the term $\propto p_3$ reads
\begin{equation}
p_3(\delta r,\delta\theta)\int_{-\pi}^{\pi}
\frac{d\varphi}{(\rho^2+P_{\varphi\varphi}\varphi^2)^{3/2}}=
\frac{2\pi p_3(\delta r,\delta\theta)}{\rho^2\sqrt{\rho_0^2+\rho^2}}\to 0
\end{equation}
as $\rho\to 0$.
The contribution from the term $\propto p_1$ also vanishes at the limit $\rho\to 0$:
\begin{eqnarray}\label{reg2}
p_1(\delta r,\delta\theta)\int_{-\pi}^{\pi}
\frac{\varphi^2\,d\varphi}{(\rho^2+P_{\varphi\varphi}\varphi^2)^{3/2}}&=&
2p_1(\delta r,\delta\theta)P_{\varphi\varphi}^{-3/2}
\left[\ln\left(\frac{\rho_0+\sqrt{\rho_0^2+\rho^2}}{\rho}\right)
-\rho_0(\rho_0^2+\rho^2)^{-1/2}\right] \nonumber\\
&=&
2P_{\varphi\varphi}^{-3/2}p_1(\delta r,\delta\theta)\times\left[-\ln(\rho/\rho_0)+O(\rho)\right]
\to 0.
\end{eqnarray}
Hence, the integral in Eq.\ (\ref{reg1}) vanishes as $\rho\to 0$ and is therefore a $C^0$ function of $\delta r$ and $\delta\theta$.

The above verifies that the modes $\Phi_{\rm R}^m$ are each continuous at the location of the particle (and elsewhere). We do not expect, however, the {\it derivatives of} $\Phi_{\rm R}^m$ to be continuous. [That the derivative are likely to be discontinuous is suggested, for example, by the form of the contribution evaluated in Eq.\ (\ref{reg2}).] In the numerical scheme to be developed in Sec.\ \ref{Sec:V} we shall assume explicitly that the solutions $\Phi_{\rm R}^m$ are continuous.

\subsection{Expressions for the puncture modes $\Phi_{\rm P}^m$}
To implement the puncture scheme set out above [Eq.\ (\ref{scheme})], we need explicit expressions for the puncture modes $\Phi_{\rm P}^m$ and for the regularized source modes $S_{\rm R}^m$. We start by obtaining the necessary expressions for $\Phi_{\rm P}^m$.

With the above choice of $\Phi_{\rm P}$ [Eqs.\ (\ref{puncturefn})--(\ref{rho})], the modes $\Phi_{\rm P}^m$ are given by
\begin{eqnarray}\label{PhiPm0}
\Phi_{\rm P}^m
&=&
\frac{q}{2\pi}
\int^{\pi}_{-\pi}\frac{e^{-im\varphi}}
{\sqrt{\rho^2 + 2P_{\varphi\varphi}(1-\cos\delta\varphi)}}\, d\varphi \nonumber\\
&=&
\frac{q}{2\pi}\,e^{-im\omega t_{\rm p}}
\int^{\pi-\omega t_{\rm p}}_{-\pi-\omega t_{\rm p}}\,\frac{e^{-imx}}
{\sqrt{\rho^2 + 2P_{\varphi\varphi}(1-\cos x)}}\, dx
\nonumber\\
&=&
\frac{q}{2\pi}\,e^{-im\omega t_{\rm p}}
\int^{\pi}_{-\pi}\,\frac{\cos(mx)}
{\sqrt{\rho^2 + 2P_{\varphi\varphi}(1-\cos x)}}\, dx ,
\end{eqnarray}
where in the second integral we have changed the integration variable as $\varphi\to x=\delta\varphi=\varphi-\omega t_{\rm p}$, and where in the third integral we (i) shifted both integration limits by $\omega t_{\rm p}$ (noticing the integrand is periodic with period $2\pi$), and (ii) made use of the fact that the imaginary part vanishes by symmetry.  For all $m=0,1,2,\ldots$, the last integral can be represented in terms of complete elliptic integrals. We find
\begin{equation}\label{PhiPm}
\Phi_{\rm P}^m =
\frac{q\,e^{-im\omega t_{\rm p}}}{2\pi P_{\varphi\varphi}^{1/2}}
\left[p^m_K(\rho) \gamma K(\gamma)+p^m_E(\rho) \gamma E(\gamma)
\right],
\end{equation}
where
\begin{equation}\label{gamma}
\gamma\equiv [1+\rho^2/(4P_{\varphi\varphi})]^{-1/2},
\end{equation}
$\tilde{K}(\gamma)$ and $\tilde{E}(\gamma)$ are complete elliptic integrals of the first and second kinds, respectively (as defined in \cite{GR1980}), and $p^m_K$ and $p^m_E$ are certain polynomials in $\rho^2$.  We tabulate these polynomials in Appendix \ref{AppA} for $m=0$--$5$.

\subsection{Expressions for the source modes $S_{\rm R}^m$}
Within our puncture scheme, the source for the field $\Phi_{\rm R}$ inside $\cal T$ is $S_{\rm R}\equiv S-\Box\Phi_{\rm P}$, and its $m$-modes are given by
\begin{equation}\label{SRm1}
S_{\rm R}^m(r,\theta;r_0)=\frac{q}{2\pi}\int_{-\pi}^{\pi}\left(S-\Box\Phi_{\rm P}\right)
e^{-im\varphi}d\varphi.
\end{equation}
With the function $\Phi_{\rm P}$ defined above, and using $\varphi=\delta\varphi+\omega t_{\rm p}$, this takes the form
\begin{equation}\label{SRm2}
S_{\rm R}^m=\frac{q}{2\pi}e^{-im\omega t_{\rm p}}\left(
S_1 I_1^m+S_2 I_2^m+S_3I_3^m+S_4I_4^m\right),
\end{equation}
where the $I^m_n$ (depending on $r_0$ only) are the integrals
\begin{eqnarray}
I_1^m&=&\int_{-\pi}^{\pi}\epsilon_{\rm
P}^{-3/2}e^{-im\delta\varphi}d(\delta\varphi),
\nonumber\\
I_2^m&=&\int_{-\pi}^{\pi}\epsilon_{\rm P}^{-3/2}\cos\delta\varphi\, e^{-im\delta\varphi}
d(\delta\varphi),
\nonumber\\
I_3^m&=&\int_{-\pi}^{\pi}\epsilon_{\rm P}^{-5/2} e^{-im\delta\varphi}
d(\delta\varphi),
\nonumber\\
I_4^m&=&\int_{-\pi}^{\pi}\epsilon_{\rm P}^{-5/2}\sin^2\delta\varphi\, e^{-im\delta\varphi}
d(\delta\varphi),
\end{eqnarray}
and where the $S_n$ are $m$-independent functions of $r$ and $\theta$ (as well as $r_0$), given by
\begin{eqnarray}
S_1&=&P_{rr}f(r)+2r^{-2}P_{rr}(r-M)\delta r+r^{-2}P_{\theta\theta}(1+\delta\theta\cot\theta),
\nonumber\\
S_2&=&P_{\varphi\varphi}\left[r^{-2}\sin^{-2}\theta-\omega^2/f(r)\right],
\nonumber\\
S_3&=&-3P_{rr}^2 f(r)\delta r^2-3r^{-2}P_{\theta\theta}^2\delta\theta^2,
\nonumber\\
S_4&=&-3P_{\varphi\varphi}^2\left[r^{-2}\sin^{-2}\theta-\omega^2/f(r)\right].
\end{eqnarray}
In obtaining Eq.\ (\ref{SRm2}) from Eq.\ (\ref{SRm1}) one should note the following:  Firstly, the function $\Phi_{\rm P}$ depends on $t$, through $\delta\varphi= \varphi-\varphi_{\rm p}=\varphi-\omega t_{\rm p}=\varphi-\omega t$ (as in our construction we take $t=t_{\rm p}$); this should be be taken into account properly when evaluating $\Box\Phi_{\rm P}$.  Secondly, the source $S_{\rm R}$ contains no distributional component (i.e., the delta functions in $S$ and $\Box\Phi_{\rm P}$ ``cancel each other'' precisely)\footnote{This can be seen by considering the volume integral of $\Box\Phi_{\rm R}$ over a small 3-ball (in $\Sigma$) surrounding the particle, at the limit where the radius of the ball tends to zero. Using the Gauss theorem, this can be converted to a surface integral of $\Phi_{{\rm R},\alpha}$ over the 2-sphere. By virtue of Eq.\ (\ref{PhiRasy}) we have that $\Phi_{\rm R}$ is bounded at the particle, and that the gradient $\Phi_{{\rm R},\alpha}$ can at most diverge as $\sim 1/\epsilon_0$ there. Hence, the surface integral of $\Phi_{{\rm R},\alpha}$ vanishes as the radius of the $2$-sphere is taken to
zero, implying $\Box\Phi_{\rm R}$ (and hence also $S_{\rm R}$) contains no Dirac deltas.}.

The integrals $I^m_{1,\ldots,4}$ can once again be expressed in terms of complete elliptic integrals. Introducing the dimensionless ``local distance in the $r$--$\theta$ plane'',
\begin{equation}
\tilde\rho\equiv \frac{\rho}{2P_{\varphi\varphi}^{1/2}}
\end{equation}
[in terms of which the quantity $\gamma$ of Eq.\ (\ref{gamma}) reads $\gamma=(1+\tilde\rho^2)^{-1/2}$], we have
\begin{eqnarray}\label{Im}
I_{n=1,2}^m&=&
P_{\varphi\varphi}^{-3/2}\gamma
\left[p^m_{nK}(\tilde\rho) K(\gamma)+\tilde\rho^{-2}p^m_{nE}(\tilde\rho) E(\gamma)\right],
\nonumber\\
I_{3}^m&=&
P_{\varphi\varphi}^{-5/2}\gamma^3
\tilde\rho^{-2}\left[p^m_{3K}(\tilde\rho) K(\gamma)+\tilde\rho^{-2}p^m_{3E}(\tilde\rho) E(\gamma)\right],
\nonumber\\
I_{4}^m&=&
P_{\varphi\varphi}^{-5/2}\gamma
\left[p^m_{4K}(\tilde\rho) K(\gamma)+\tilde\rho^{-2}p^m_{nE}(\tilde\rho) E(\gamma)\right],
\end{eqnarray}
where $p^m_{nK}$ and $p^m_{nE}$ are polynomials in $\tilde\rho^2$, which are all non-zero at $\tilde\rho\to 0$. (Note the particle limit corresponds to $\tilde\rho\to 0^+$, or $\gamma\to 1^-$.] The polynomials $p^m_{nK}$ and $p^m_{nK}$ are tabulated in Appendix \ref{AppA} for $m=0$--$5$.

\subsection{Behavior of $S^m_{\rm R}$ near the worldline}\label{localexp}
Even though the modes $\Phi_{\rm R}^m$ are finite (and continuous) at the worldline, the source modes $S_{\rm R}^m$ can still diverge there. Indeed, as we show below, $S_{\rm R}^m$ diverges like $\rho^{-1}$ as $\rho\to 0$ (with coefficient that depends on the direction of approach in
the $r$--$\theta$ plain). This is a serious concern when it comes to numerical implementation, since the finite-difference scheme would normally require to evaluate the source $S_{\rm R}^m$ also at $\rho=0$, where it diverges. We will deal with this complication by integrating ``by hand'' numerical grid points which lie on the worldline (this procedure will be described in Sec.\ \ref{Sec:V}). For this, we shall need some information on the asymptotic form of $S_{\rm R}^m$ near $\rho=0$. It will prove necessary to have at hand the form of $S_{\rm R}^m$ up to $\mathcal{O}(\rho^0)$ (inclusive). We now derive the necessary asymptotic formula.

Consider the form of $S_{\rm R}^m$ as given in Eq.\ (\ref{SRm2}). The functions $S_n$ are easily expanded in powers of $\delta r$ and $\delta\theta$. We wish to rewrite this expansion in terms of local polar coordinates as in Eq.\ (\ref{polars}). However, our numerical coordinates will be based on $t,r_*$ rather than $t,r$, and it will prove advantageous to replace $\rho$ as our local polar coordinate with a new coordinate, based on $\delta r_*\equiv r_*-r_{*0}$. We hence introduce the new local polar coordinates $\rho_*,\phi_*$, defined by
\begin{eqnarray}\label{rhostar}
\delta r_*   &=& \rho_{*}f_0^{-\frac{1}{2}} \cos\phi_*, \nonumber\\
\delta\theta &=& \rho_{*}r_0^{-1} \sin\phi_*.
\end{eqnarray}
(The coordinates $\rho_*,\phi_*$ coincide with $\rho,\phi$ at leading order in $\rho$, but deviate at higher order.) Using $\delta r = f_0 \delta r_* + \frac{1}{2}f'(r_0)\delta r_*^2 + \cdots$ and Eq.\ (\ref{rhostar}) we can then express each of the $S_n$ as an expansion in $\rho_*$.  To expand the $I_n^m$ in powers of $\rho_*$, we first expand the Elliptic functions in Eq.\ (\ref{Im}) in powers of $\rho$, using
\begin{eqnarray}
K(\gamma)&=&-\ln(\chi/4)-\frac{1}{4}\chi^2[\ln(\chi/4)+1]+\mathcal{O}(\chi^4\ln\chi),
\nonumber\\
E(\gamma)&=&1-\frac{1}{4}\chi^2[2\ln(\chi/4)+1]+\mathcal{O}(\chi^4\ln\chi),
\end{eqnarray}
[Eqs.\ (8.113-3) and (8.114-3) of Ref.\ \cite{GR1980}]
where $\chi\equiv\sqrt{1-\gamma^2}=\tilde\rho(1+\tilde\rho^2)^{-1/2}$
(recall the particle limit corresponds to $\tilde\rho\to 0^+$, or $\gamma\to 1^-$, and so also to $\chi\to 0^+$).  We then re-expand $I_n^m(\rho)$ in powers of $\rho_*$ using
\begin{equation}\label{rhostarexp}
\rho=\rho_{*} + \left[\frac{M}{r_0^{2}f_0^{3/2}} \cos^3\phi_*\right]\rho_{*}^2
+\left[ \frac{M(11M-4r_0)}{6r_0^4f_0^3}\cos^4\phi_*\sin^2\phi_*
-\frac{2M}{3r_0^3f_0^2}\cos^6\phi_*\right]\rho_{*}^3
+\mathcal{O}(\rho_*^4),
\end{equation}
obtained by substituting $\delta r = f_0 \delta r_* + \frac{1}{2}f'(r_0)\delta r_*^2$ in  Eq.\ (\ref{rho}), and then substituting for $\delta r_*$ and $\delta \theta$ from Eq.\ (\ref{rhostar}).
Inserting all the above expansions in Eq.\ (\ref{SRm2}), we obtain the following asymptotic formula for the source modes:
\begin{equation}\label{SRmLocal}
S_{\rm R}^m = \frac{q}{2\pi}\,e^{-im\omega t_{\rm p}}
\left[\frac{\alpha(\phi_*)}{\rho_*} +
\beta^m_{\rm ln}\ln(\tilde\rho_*/4)+
\beta^m(\phi_*)\right] + \mathcal{O}(\rho_* \ln\tilde\rho_*),
\end{equation}
where $\tilde\rho_*\equiv\rho_*/(2P_{\varphi\varphi}^{1/2})$, and the coefficients are given by
\begin{eqnarray}\label{alpha}
\alpha(\phi_*) &=& \frac{8(1-M/r_0)}{r_0^2{\cal E}}\cos\phi_*\sin^2\phi_*,\\
\beta_{\rm ln} &=& -\frac{1}{2 P_{\varphi\varphi}^{3/2}} \label{betaln}\\
\beta^m(\phi_*) &=& \beta^m_0 + \beta_1 \cos 2\phi_* + \beta_2\cos 4\phi_* +
\beta_3 \cos 6\phi_*. \label{beta}
\end{eqnarray}
In the last expression $\beta_{n}^m$ are constant coefficients (depending on $r_0$ and $m$ only), of which only $\beta_0^m$ will be needed in what follows---this coefficient is given explicitly in Appendix \ref{AppB} for $m=0$--$5$ .

We note the following:
(i) At leading order we have $S_{\rm R}^m\propto \rho_*^{-1}$, with coefficient that
depends on $m$ only through the trivial factor $e^{-im\omega t_{\rm p}}$.
(ii) The leading-order divergence of $S_{\rm R}^m$ is direction-dependent
(it depends on the azimuthal angle $\phi_*$ in the $r$--$\theta$ plane).
(iii) Upon averaging over directions, the $\sim\rho_*^{-1}$ divergence of
$S_{\rm R}^m$ cancels out; the direction-averaged singularity of $S_{\rm R}^m$
is only $\propto\ln\rho_*$. 
We will make good use of this latter observation below, when setting out our numerical scheme.

\section{Numerical Implementation: Vacuum case}\label{Sec:IV}
In this section we develop our 2+1-D numerical evolution code and test it for vacuum perturbations. To this end we set, for now, $S^m=0$ in the field equation (\ref{psis}), and consider the vacuum evolution of prescribed initial data. For each mode $m\ge 0$ we discretize the field equation using a 2nd-order-convergent finite-difference scheme, on a fixed 2+1-D mesh which is based on mixed characteristic and spatial coordinates. We test the validity of the code by (i) demonstrating 2nd-order convergence, (ii) examining the late-time decay pattern of compact initial perturbations, and (iii) comparing the numerical solutions to those obtained using evolution in 1+1-D.

\subsection{Numerical domain}
\begin{figure}[htb]
\includegraphics[width=6cm]{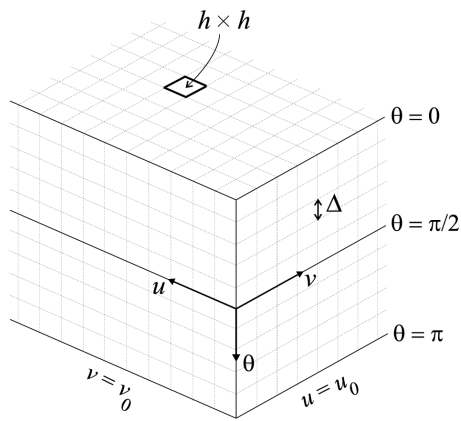}
\caption{A diagram illustrating the 2+1-D numerical domain.
The grid is based on characteristic (Eddington-Finkelstein) coordinates $u$ and $v$ and Schwarzschild coordinate $\theta$. Initial data are specified on the (null) surfaces $v=0$ and $u=0$. Boundary conditions are specified at the ``poles'', $\theta=0,\pi$.}
\label{numericalgrid}
\end{figure}
The 2+1-D numerical domain consists of a ``stack'' of staggered double-null 1+1-D grids, each based on $u,v$ coordinates---see Fig.\ \ref{numericalgrid}. We denote the grid spacing in each of $u,v$ by $h$, and the grid spacing in $\theta$ by $\Delta$. The evolution starts with initial data on the two hypersurfaces defined by $v=v_0$ and $u=u_0$. (In the circular-orbit case considered in the next section, these will be taken such that the initial  vertex $u_0,v_0$ corresponds to $r=r_0$, $t=0$, where $r=r_0$ is the orbital radius.) The numerical evolution proceeds first along
$\theta$, then along $u$ and finally along $v$. That is, for each $v$ value we solve for all $u$, and for each $u,v$ values we solve for all $\theta$. Boundary conditions (see below) are placed along the two surfaces $\theta=0,\pi$, representing the two polar axes.

As the grid is not based on purely characteristic coordinates, we must constrain the relation between $h$ and $\Delta$. On theoretical grounds, for the scheme to be stable it is necessary that the numerical domain of dependence contains the physical, continuum domain of dependence at each point in the evolution (``Courant condition''; see, e.g., \cite{NumRec}).  In our scheme, a grid point at $(t,\theta)$ will effectively require data from points $(t - h, \theta\pm\Delta)$, so the above condition translates to $\Delta/h \geq f^{1/2}/r$. The function $f^{1/2}/r$ attains a maximum value of $\sim 0.19245\, M^{-1}$ (at $r=3M$), so to make sure that the Courant condition
is met everywhere, we shall always take $\Delta/h \geq 0.2\,M^{-1}$.

\subsection{Finite difference scheme}
\begin{figure}[htb]
\includegraphics[width=6cm]{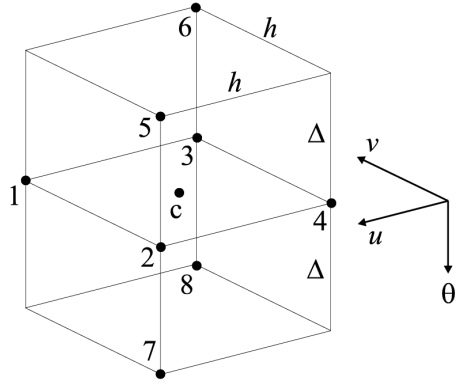}
\caption{A single numerical grid cell, of coordinate dimensions $h\times h\times (2\Delta$).
Finite difference approximations are made about the point c in terms of points 1--8. The equations are then rearranged to give an evolution scheme for the field at point 1, based on the values at points 2--8 solved for in previous steps of the evolution.}
\label{numericalcell}
\end{figure}
Our numerical evolution scheme is constructed from finite-difference approximations to the terms in Eq.\ (\ref{psis}), centred on the point c as shown in Fig.\ \ref{numericalcell}. At each stage we solve for the point 1 based on information from the points 2--8. We then rearrange the resulting formula to obtain an evolution scheme for the point 1. The various terms in Eq.\ (\ref{psis}) are approximated at point c using the centred formulas
\begin{mathletters}
\begin{equation}\label{uv0}
\Psi^m_{{\rm c},uv} = \frac{\Psi^m_1 + \Psi^m_4 -\Psi^m_3 -\Psi^m_2 }{h^2} +
\mathcal{O}(h^2),
\end{equation}
\begin{equation}\label{thetatheta0}
\Psi^m_{{\rm c},\theta\theta} =
\frac{\Psi^m_5
+ \Psi^m_6 + \Psi^m_7 + \Psi^m_8 - 2(\Psi^m_2 + \Psi^m_3)}{2\Delta^2}
+\mathcal{O}(\Delta^2, h^2),
\end{equation}
\begin{equation}\label{theta0}
\Psi^m_{{\rm c},\theta} = \frac{\Psi^m_5 + \Psi^m_6 - \Psi^m_7 - \Psi^m_8}{4\Delta}
+\mathcal{O}(\Delta^2,h^2),
\end{equation}
\begin{equation}\label{average21}
\Psi^{m}_{\rm c} = \frac{\Psi^{m}_2 + \Psi^{m}_3}{2} + \mathcal{O}(h^2).
\end{equation}
\end{mathletters}
All $r,\theta$-dependent coefficients in the field equation are simply evaluated at point $c$.  Solving for $\Psi^{m}_1$ we obtain our finite difference scheme for the vacuum case: 
\begin{eqnarray}\label{21findif}
\Psi^m_1&=&
\Psi^m_{2}+\Psi^m_{3}-\Psi^m_{4} +\frac{h^2f_{\rm c}}{8r_{\rm c}^2}\left[
\right.    \nonumber\\
&&
\left(\Psi^m_{5} + \Psi^m_{6} + \Psi^m_{7}
+ \Psi^m_{8} - 2\Psi^m_{2} -2\Psi^m_{3}\right)/\Delta^2   \nonumber\\
& &+\cot\theta_{\rm c}\left(\Psi^m_{5}+\Psi^m_{6}-\Psi^m_{7}-\Psi^m_{8}\right)
/(2\Delta)                                                        \nonumber\\
&&\left. -\left(2M/r_{\rm c}+m^2\csc^2\theta_{\rm c}\right)
\left(\Psi^m_{2}+\Psi^m_{3}\right)  \right]
+\mathcal{O}(h^2\Delta^2,h^4).
\end{eqnarray}
Here $r_{\rm c}$ and $\theta_{\rm c}$ are the values of $r$ and $\theta$ at point c,
and $f_{\rm c}=f(r_{\rm c})$.

For a fixed ratio $\Delta/h$, the finite-difference scheme (\ref{21findif}) has, effectively, a local discretization error of $\mathcal{O}(h^4)$, leading to a global (accumulated) error of $\mathcal{O}(h^2)$.\footnote{The relation between local and global errors in the scheme (\ref{21findif}) can be explained as follows: At each grid point, the field accumulates local errors from $\propto h^{-2}$ points belonging to the same $\theta$=const slice. [Note that the leading-order contribution to $\Psi^m_1$ in Eq.\ (\ref{21findif}) comes from points 2--4, which lie on the same $\theta$=const slice as point 1.] Assuming the local $\mathcal{O}(h^4)$ errors are not strongly correlated, they
accumulate to give a global error of $\mathcal{O} (h^{-2}\times h^4)= \mathcal{O}(h^2)$.} Hence, we expect the algorithm to exhibit quadratic point-wise convergence (for smooth initial data).

\subsection{Boundary conditions at the poles}
The boundaries in $u,v$ are null and thus are never encountered during the evolution. On the other hand, at the poles ($\theta = 0, \pi$) we require suitable boundary conditions. One can obtain the necessary conditions by imposing regularity of the field at the poles: Each azimuthal mode $m\neq 0$ has harmonic dependence on $\varphi$; continuity of the field across the poles (where $\varphi$ is indefinite) therefore implies that the field must vanish there.  The remaining, axially-symmetric, $m=0$ mode is symmetric across each pole (invariant under $\varphi\to-\varphi$), and so for the field to have continuous derivatives there, these derivatives must vanish. The physical boundary conditions at the poles are therefore
\begin{equation}
\Psi^{m\ne 0}(\theta=0,\pi)=0, \quad \quad \Psi_{,\theta}^{m=0}(\theta=0,\pi)=0.
\end{equation}

To implement these conditions in our code, we simply set $\Psi=0$ at the poles for all $m\ne 0$, whereas for $m=0$ we use the extrapolation
\begin{eqnarray} \label{BC}
\Psi^{m=0}(\theta=0)&=&
\frac{1}{3} \left[4\Psi^{m=0}(\theta=\Delta)-\Psi^{m=0}(\theta=2\Delta)\right] +
\mathcal{O}(\Delta^4), \nonumber\\
\Psi^{m=0}(\theta=\pi)&=&
\frac{1}{3} \left[4\Psi^{m=0}(\theta=\pi-\Delta)-\Psi^{m=0}(\theta=\pi-2\Delta)\right] +
\mathcal{O}(\Delta^4).
\end{eqnarray}
[Here the error term is $\mathcal{O}(\Delta^4)$, rather than $\mathcal{O}(\Delta^3)$, since $\Psi^{m=0}$ is an even function of $\theta$ at the poles.]

\subsection{Tests of vacuum code}
For the following tests we specified initial data in the form an ``outgoing'' narrow pulse starting at $v_0,u_0$, with a certain $\theta$-profile chosen differently for each of the tests (see below). In all cases we took $v_0=r_*(r=7M)$ and $u_0=-r_*(r=7M)$. We selected $\Delta$ such that, at the lowest resolution, $\pi/\Delta$ is an integer number (and so an integer number of $\Delta$ intervals fit into our grid between the two poles). In all cases we fixed the ratio $\Delta/h$ at $2\pi/5\, M^{-1}\sim 1.26\, M^{-1}$. This is safely above the Courant limit, and for our lowest resolution ($h=M/4$) gave sufficient $\theta$-resolution ($\Delta=\pi/10$) to resolve the lowest few multipoles.

\subsubsection{Numerical convergence}
We tested the point-wise self-convergence rate of the above scheme by examining the solutions along various 1-D cross sections of the 2+1-D grid, for a geometrical sequence of decreasing $h$ values approaching $h\to 0$. Specifically, we looked at $\Psi^m(t)$ along $(r,\theta)=(7M,\pi/2)$ and at $\Psi^m(\theta)$ along $(t,r)=(400M,7M)$, for resolutions $h=M/2$, $M/4$, and $M/8$. For initial data, we took a narrow distribution, centered at $(v,u,\theta)=(v_0,u_0,\pi/2)$. We deliberately chose a discontinuous initial distribution (a narrow square pulse), which simulates the
situation in the particle case (see below) and allows us to assess the effect of non-smoothness in the initial data on the convergence rate.

\begin{figure}[htb]
\includegraphics[width=0.49\textwidth]{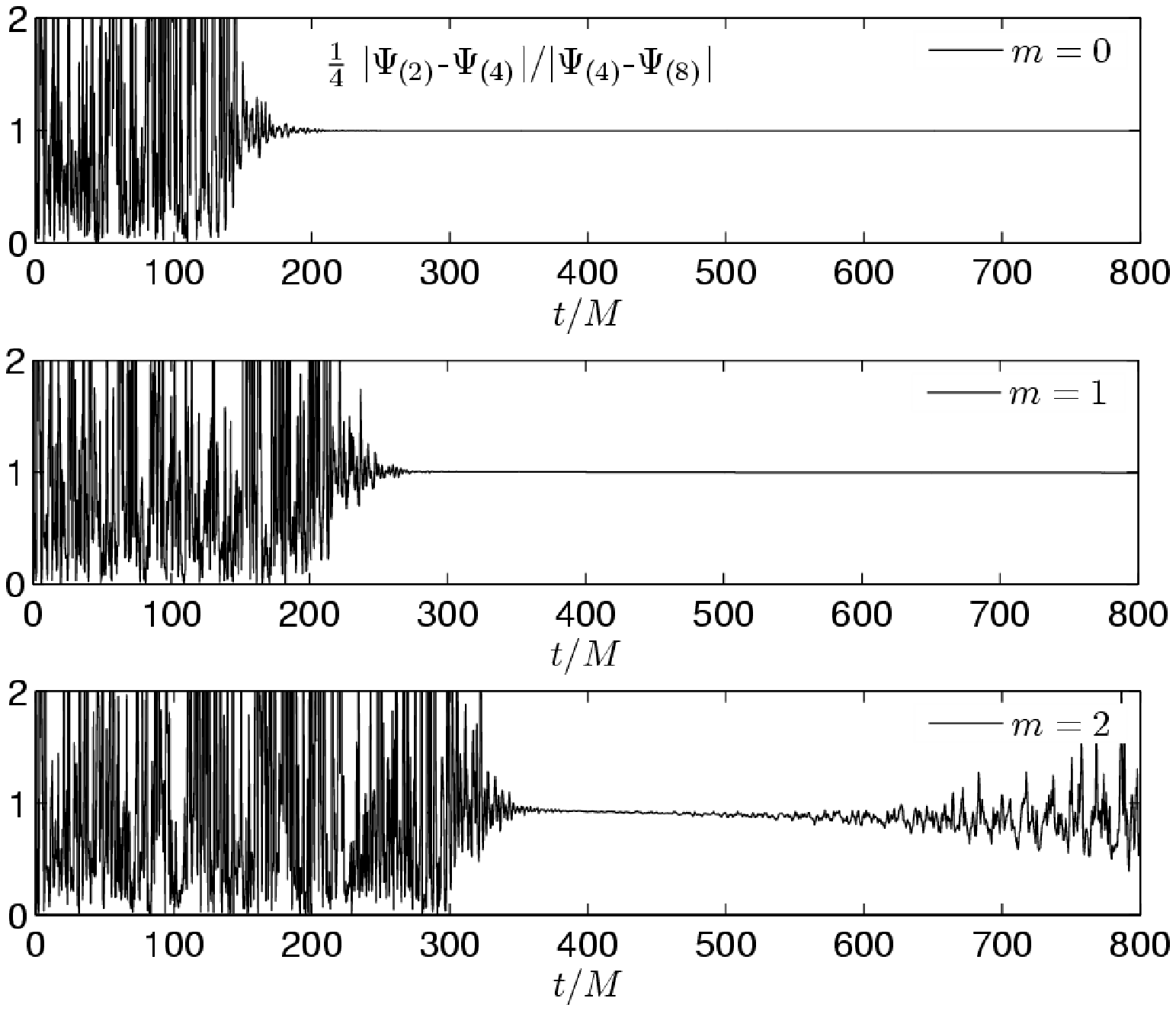}
\includegraphics[width=0.49\textwidth]{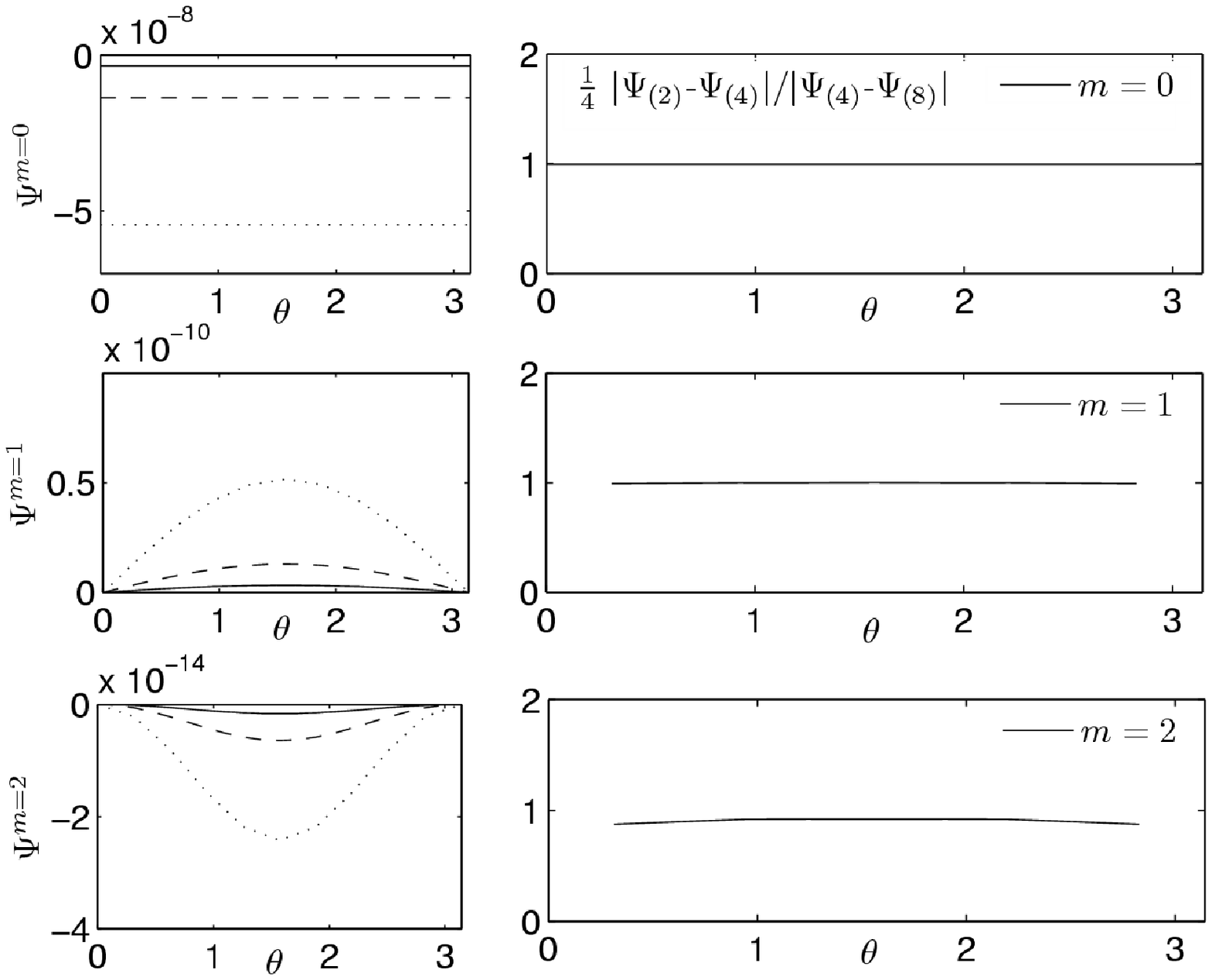}
\caption{Numerical convergence test for vacuum perturbations. Upper, middle and lower panels correspond to $m=0,1,2$, respectively. Each of the 6 panels labeled `$m=\ldots$' displays $\frac{1}{4}\times$ the relative difference $\delta\Psi_{\rm rel}^m\equiv\left|(\Psi^m_{(2)}-\Psi^m_{(4)})/(\Psi^m_{(4)}-\Psi^m_{(8)})\right|$, where $\Psi^m_{(n)}$ is the solution obtained with resolution $h=[5M/(2\pi)]\Delta=M/n$. A value of unity indicates quadratic convergence. Left panels show $\frac{1}{4}\delta\Psi_{\rm rel}^{m}(t)$ along $(r,\theta)=(7M,\pi/2)$; right panels show $\frac{1}{4}\delta\Psi_{\rm rel}^{m}(\theta)$ at $(t,r)=(400M,7M)$. For $\delta\Psi_{\rm rel}^{m}(\theta)$ we also show the solutions $\Psi^m_{(n)}$ themselves (3 small panels): $\Psi^m_{(2)}$ in dotted line,  $\Psi^m_{(4)}$ in dashed line, and $\Psi^m_{(8)}$ in solid line. These serve to demonstrate how the late-time solutions are dominated by the lowest allowed $\ell$-mode for a given $m$, i.e.,
$\ell_{\rm min}=|m|$. The solutions show good 2nd-order numerical convergence at late time.
At the early stage of the evolution, the solutions are affected by the initial-data discontinuity, bouncing back and forth between the two poles; however, this effect gradually dies off over time through dissipation. The noise in $\delta\Psi_{\rm rel}^{m=2}$ at very late time is due to round-off
truncation error, which kicks in when the amplitude of $\Psi^{m=2}$ drops very low.}
\label{convergence}
\end{figure}

As demonstrated in Fig.\ \ref{convergence}, the vacuum numerical evolution shows a clear second-order point-wise convergence at late time. Early in the evolution, multiple reflection of the discontinuous data off the $\theta$-boundaries introduces large numerical error [for grid cells in which $\Psi$ is discontinuous, the finite-difference formula (\ref{21findif}) has a local error of $\mathcal{O}(h^2)$ rather than $\mathcal{O}(h^4)$]. However, over time the discontinuity dissipates, and quadratic convergence is retained.

\subsubsection{Late time tails}
In theory, after the initial burst of radiation and ringing phase, the field should settle down to a power-law decay at late time. The exponent of this decay is determined predominantly by the lowest multipole number $\ell$ present in the data, but is otherwise independent of the shape of the initial data . In the case of a Schwarzschild background (where different $\ell$-modes do not couple) and compact initial data with angular dependence of a pure $\ell$-mode, we expect
late-time tails of the form $\Psi^m\propto t^{-2\ell-3}$ at fixed $r$ (i.e., for $t\gg M$ with $t\gg r$), and of the form $\Psi^m\propto u^{-\ell-2}$ at null infinity (i.e., for $v\gg M$ with $v\gg u$) \cite{P1972}.

\begin{figure}[htb]
\includegraphics[width=0.49\textwidth]{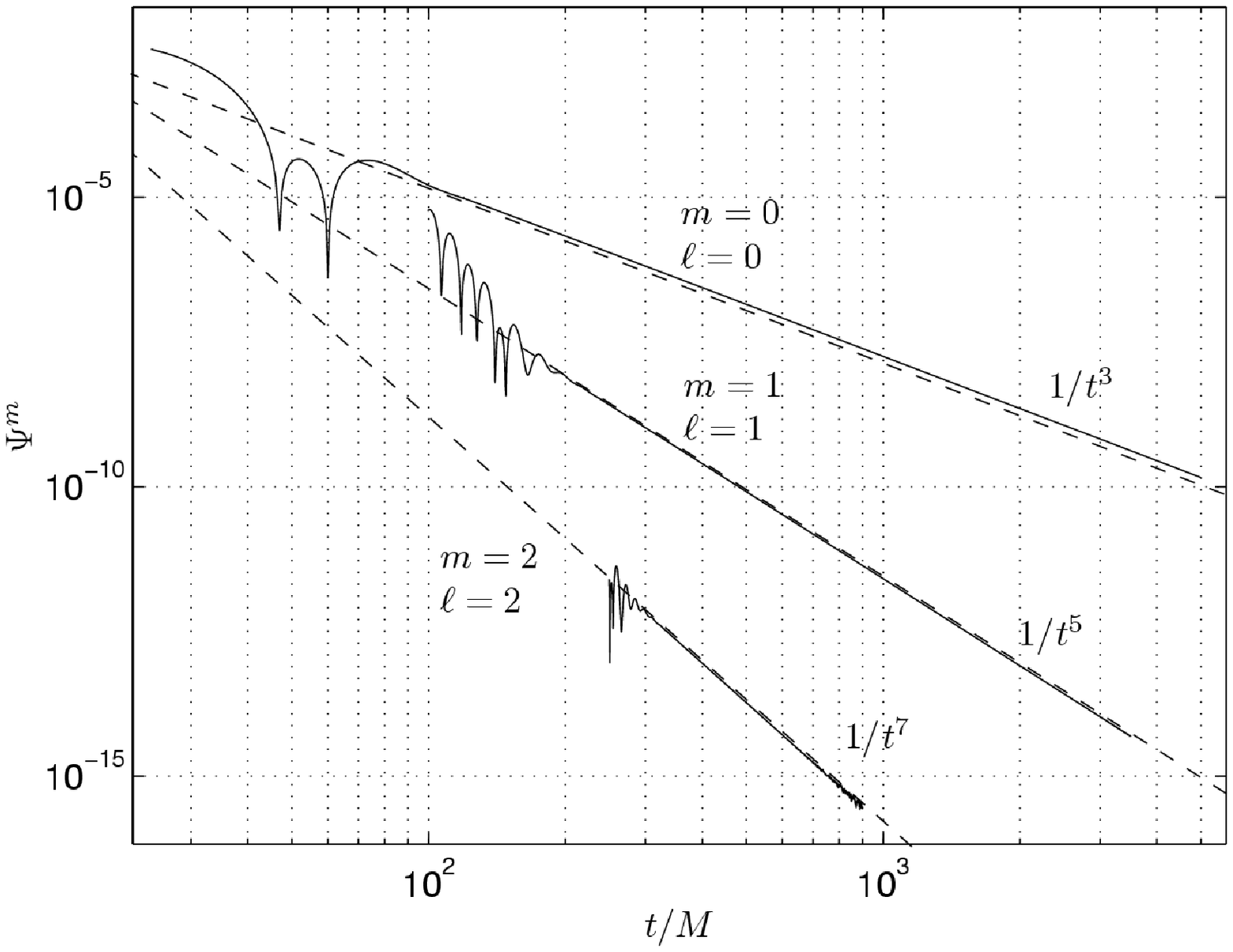}
\includegraphics[width=0.49\textwidth]{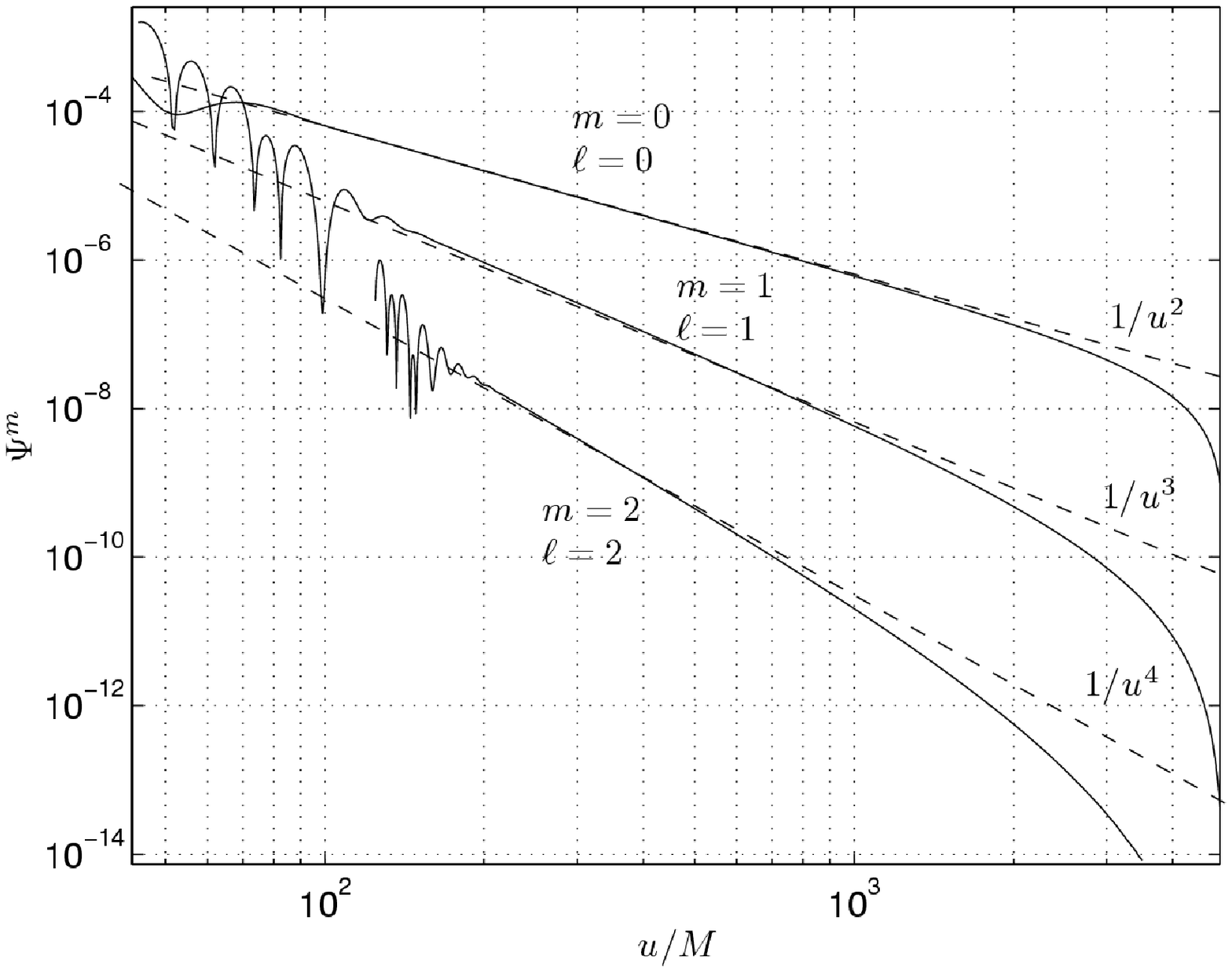}
\caption{
Late-time power-law decay tails of $\Psi^{m}$, with initial perturbation made of a single, pure $\ell=m$ mode. Left panel (solid lines): tails along lines of constant $r=7M, \theta=\pi/2$. Right panel (solid lines): tails at ``null infinity'', read off along $v=\mathrm{const}=1000M$, again at $\theta=\pi/2$. (For clarity, part of the ringing phase data has been removed in these figures.) The dashed lines are reference lines $\propto t^{-2\ell-3}$ (left panel) and $\propto u^{-\ell-2}$ (right panel), showing the theoretical asymptotic slopes. The gradual deviation from the predicted slopes in the ``null infinity'' data is explained by the fact that the large-$u$ regime ($u$ comparable to $v$) no longer approximates null infinity.
}\label{latetimetails}
\end{figure}

To test these predictions with our vacuum code, we specified initial data in the form of a compact outgoing pulse of a pure $\ell$-mode content. Specifically, we took $\Psi^m(u=u_0)=0$ for all $v,\theta$, and $\Psi^m(v=v_0)=\sin^2[\pi (u-u_0)/(8M)]P_{\ell m}(\cos\theta)$ for $0\leq u\leq 8M$,  with $\Psi^m(v=v_0)=0$ for $u>8M$. Here $P_{\ell m}(\cos\theta)$ is the associated Legendre polynomial. The left panel in Fig.\ \ref{latetimetails} shows the late-time decay tails of our vacuum solutions at fixed $r(=7M$), for $\ell=m=0,\,1,\,2$. The right panel in Fig.\ \ref{latetimetails} shows the decay tails as a function of retarded time $u$ at large $v(=1000M)$, approximating null infinity. The decay rates are in excellent agreement with the theoretical prediction.

\subsubsection{Comparison with 1+1-D solutions}\label{comp1+1d}
The best quantitative test of our code comes from comparison with results obtained using an independent evolution code formulated in 1+1-D (time+radius).  For this comparison we wrote a 1+1-D code similar to the one developed in Ref.\ \cite{BB2000}. In the 1+1-D treatment we construct $\Phi$ through an expansion in spherical harmonics, $\Phi=(1/r)\sum_{\ell m}\Psi^{\ell m}(t,r)Y^{\ell m}(\theta,\varphi)$, where the time-radial functions $\Psi^{\ell m}(t,r)$ are obtained using characteristic evolution in 1+1-D (see \cite{BB2000} for details). Suppose that $\Psi^{\ell m}_{1+1}$ is a 1+1-D solution for given $\ell,m$ and for initial data in the form of a compact outgoing pulse with some profile $U(u)$. Suppose also that $\Psi^{m}_{2+1}$ is a 2+1-D solution with the same $m$, for initial data in the form of a compact outgoing pulse with a profile $U(u)P_{\ell m}(\cos\theta)$. Then, we expect the solutions to be related by $\Psi^{m}_{2+1}(t,r,\theta)=a_{\ell m}\Psi^{\ell m}_{1+1}(t,r)P_{\ell m}(\cos\theta)$, where $a_{\ell m}$ are the normalization coefficients appearing in the relation between the spherical harmonics and the Legendre polynomials: $Y^{\ell m}=a_{\ell m}P_{\ell m}(\cos\theta)e^{im\varphi}$.

\begin{figure}[htb]
\includegraphics[width=0.49\textwidth]{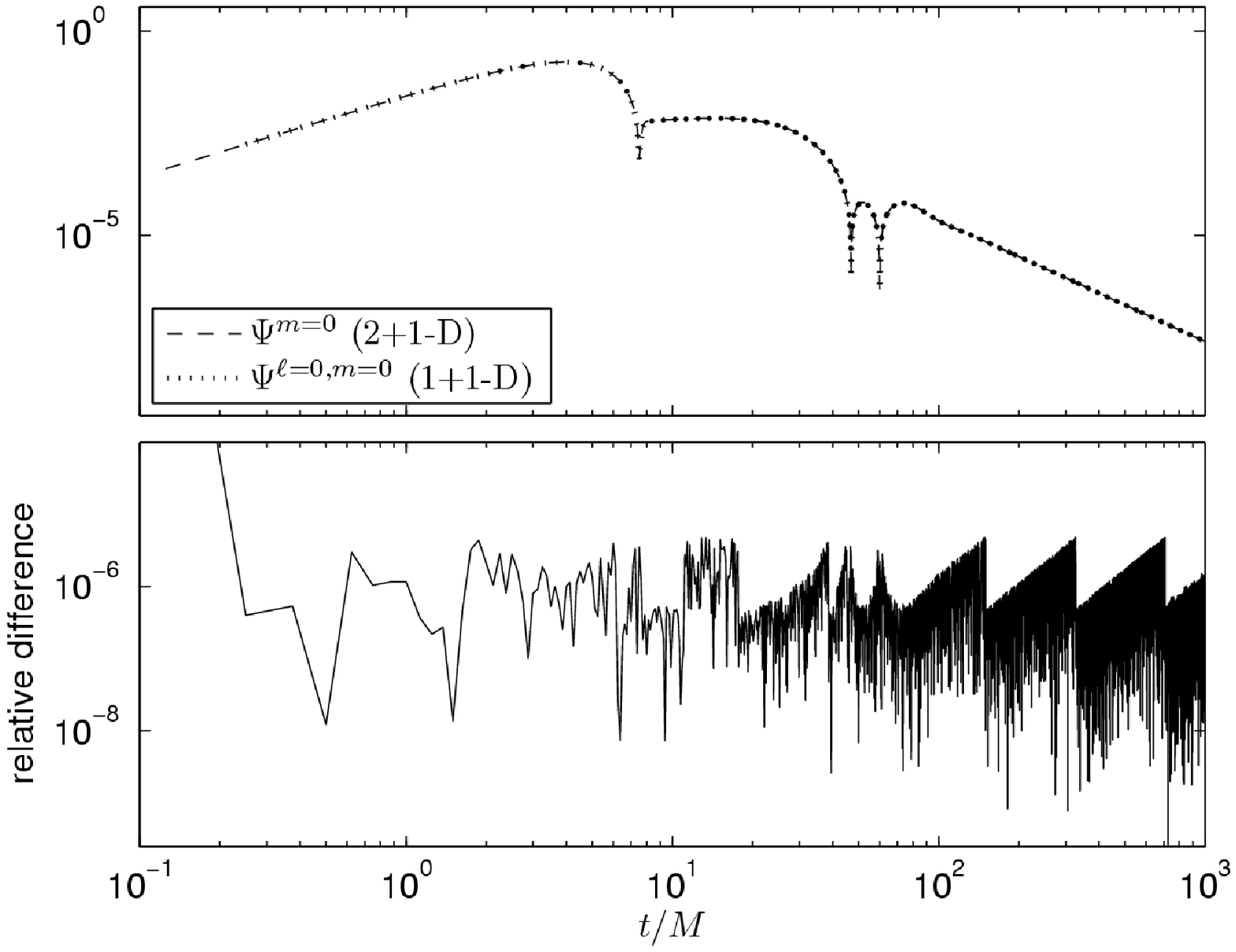}
\includegraphics[width=0.49\textwidth]{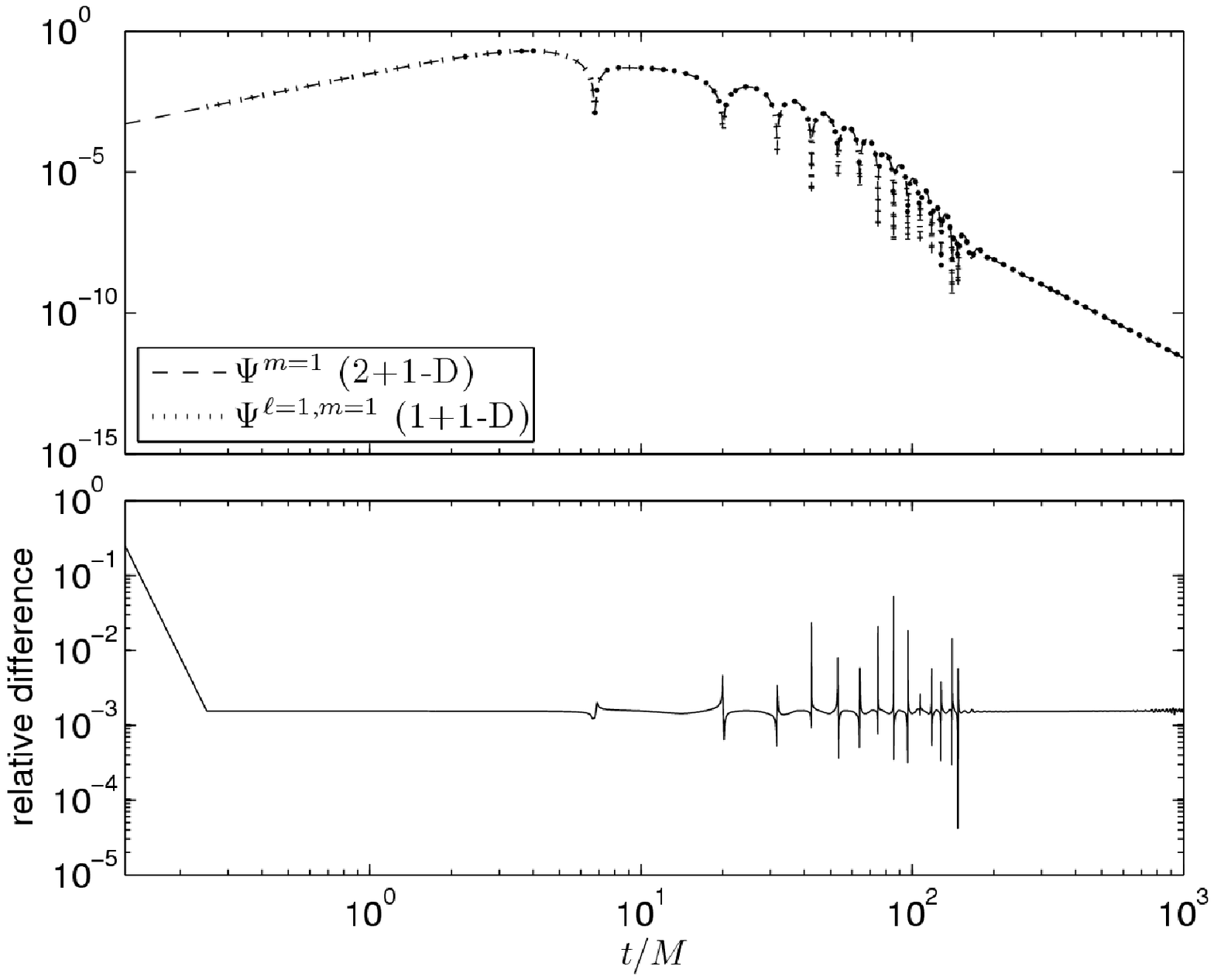}
\caption{
Comparison between vacuum solutions obtained with our 2+1-D code, and solutions obtained independently using 1+1-D evolution. Both solutions correspond to the same physical initial data, containing a single, pure $\ell,m$ mode (see text for details). We compare them here as functions of $t$, at fixed $r=7M$ and $\theta=\pi/2$. The left and right panels display $\ell=m=0$ and $\ell=m=1$, respectively. The upper panels show, superposed, both $\Psi^m_{1+1}(t)$ (dotted line) and $\Psi^m_{2+1}(t)$ (dashed line). The lower panels show the relative differences $2\left|(\Psi^m_{1+2}-\Psi^m_{1+1})/(\Psi^m_{1+2}+\Psi^m_{1+1})\right|$. The small relative difference is (presumably) due to finite-differentiation errors in both codes, which in the 2+1-D code also include a small amount of ``contamination'' from coupling to higher $\ell$-modes. (For $\ell=m=0$ the tiny relative difference is dominated by noisy round-off error.) The good agreement between the 2+1-D
and 1+1-D solutions provides a strong validation test for the 2+1-D code.}
\label{compare}
\end{figure}

For the comparison, we ran the 2+1 code with the same initial data as for the tail-test above. We then ran the 1+1-D code with corresponding $\ell,m$ and $u$-profile. The plots in Fig.\ \ref{compare} display results from this comparison (for $\ell=m=0$ and $\ell=m=1$, along lines of constant $r,\theta$). We find a good agreement between the 2+1-D and 1+1-D solutions.

\section{Numerical implementation: Circular orbit}\label{Sec:V}
\subsection{Inclusion of the particle; the worldtube}

\begin{figure}[htb]
\includegraphics[width=7cm]{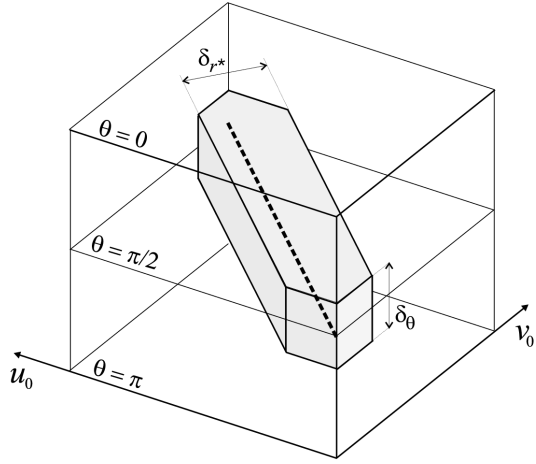}
\caption{The worldtube configuration. The sketch illustrates the geometry
of the numerical domain in the circular-orbit case. We show a portion
of the numerical grid, containing the worldline (dashed line), and
the worldtube surrounding it (shaded volume). More details are given
in the text.}
\label{fig:tube}
\end{figure}

We now come to the main part of our analysis: the inclusion of the particle
through the puncture scheme described in Sec.\ \ref{Sec:III}. We consider the
physical setup described in Sec.\ \ref{subsec:setup}, i.e., a scalar-charge
particle set in an equatorial circular geodesic orbit with radius $r=r_0$ around the
Schwarzschild black hole. The scalar field equation now has source $S$, given
in Eq.\ (\ref{source}). In our $2+1$-D numerical domain the particle traces a
straight line along $v=u+2r_{*0}$, $\theta=\pi/2$, where $r_{*0}\equiv
r_*(r_0)$ (see Fig.\ \ref{fig:tube}).
We set up the grid such that the initial vertex $(v_0,u_0)$ corresponds to
$t=0$ and $r_*=r_{*0}$; namely, we take $v_0=r_{*0}$ and $u_0=-r_{*0}$.
We select the $\theta$ grid separation $\Delta$ such that $\pi/\Delta$ is an
even integer. With this setup, which turns out most convenient, the particle
cuts straight through grid points precisely every $\Delta t=h$.

To solve the sourced evolution problem, we implement our puncture scheme
as formulated in Eq.\ (\ref{scheme}). We first introduce a ``worldtube'' $\cal T$
within the numerical grid. For convenience, we choose a worldtube with a uniform
rectangular cross section: For any fixed time $t$ we take it to be the region
$r_{*0}-\delta_{r_*}/2\leq r_*\leq r_{*0}+\delta_{r_*}/2$,
$\pi/2-\delta_{\theta}/2\leq \theta\leq \pi/2+\delta_{\theta}/2$, where the
``width'' $\delta_{r_*}$ and ``height'' $\delta_{\theta}$ of the tube are kept
as (two independent) control parameters in our analysis; See Fig.\ \ref{fig:tube}
for an illustration of the worldtube setup.
We will typically take $\delta_{r_*}$ and $r_0\delta_{\theta}$ to be of order a
few $M$. Among the robustness tests for our code, we will establish that
the numerical solutions are independent of $\delta_{r_*}$ and $\delta_{\theta}$
(up to numerical error which decreases with grid size).

\subsection{Finite-difference scheme} \label{FDS}

For our numerical treatment, we reformulate the puncture scheme (\ref{scheme})
in terms of the field variable $\Psi^m=r\Phi^m$, as in the vacuum case.
The scheme becomes
\begin{equation}\label{schemePsi}
\left\{ \begin{array}{ll}
\Box_{\Psi}^m \Psi^m_{\rm R}= -(fr/4)S^m_{\rm R}\equiv Z^m_{\rm R}
                & {\rm in\ } \cal T, \\
\Box_{\Psi}^m \Psi^m=0
                & {\rm outside\ } \cal T, \\
{\rm with \ }\Psi^m_{\rm R}= \Psi^m-r\Phi^m_{\rm P}
                & {\rm on\ } \partial \cal T, \\
\end{array}
\right.
\end{equation}
where, recall, the operator $\Box_{\Psi}^m$ is defined in Eq.\ (\ref{psis}),
and $S^m_{\rm R}$ and $\Phi^m_{\rm P}$ are given analytically in Eqs.\
(\ref{SRm2}) and (\ref{PhiPm}), respectively. The evolution algorithm is
similar to the one applied in the vacuum case: Starting with initial data on
$v=v_0$ and $u=u_0$ (see below), we integrate along ``planes'' of fixed $v$, where
on each such plane we integrate along ``lines'' of fixed $u$.
Consider again a typical, single grid cell as depicted in Fig.\ \ref{numericalcell}:
The values at points 2--8 are assumed to have been solved for in previous steps, and
we need to obtain the value at point 1. The algorithm first labels each of the points
1--8 as either `out' or `in', depending on whether it lies outside or inside $\cal T$,
respectively. If a point lies on $\partial{\cal T}$ it is labeled `in'.
Four cases are possible:
{\it Case 1:} All points 1--8 are `out'. In this case the integrator implements the vacuum scheme
(\ref{21findif}) to solve for $\Psi_{1}^m$, just like in the global vacuum case.
{\it Case 2:} All points 1--8 are `in'. The integrator then implements a different scheme, described below,
which is based on the sourced equation $\Box_{\Psi}^m \Psi^m_{\rm R}=Z^m_{\rm R}$.
{\it Case 3:} Point 1 is `out', but some of the points 2--8 are `in'. In this case the values of
the `in' points are adjusted according to $\Psi^m_{\rm R}\to\Psi^m=\Psi^m_{\rm R}+r\Phi^m_{\rm P}$,
after which the integrator solve for $\Psi_{1}^m$ using the vacuum scheme (\ref{21findif}).
{\it Case 4:} Point 1 is `in', but some of the points 2--8 are `out'. Then the values of the `out'
points are adjusted according to $\Psi^m\to \Psi^m_{\rm R}=\Psi^m-r\Phi^m_{\rm P}$,
after which the integrator obtains $\Psi^m_{\rm R}$ at point 1 using the sourced-equation
scheme described below.
This way, the algorithm effectively solves for $\Psi^m$ outside $\cal T$ and for
$\Psi_{\rm R}$ inside $\cal T$, adjusting the integration variable across the boundary
$\partial{\cal T}$.

We now describe the finite-difference scheme applied inside $\cal T$.
Referring again to Fig.\ \ref{numericalcell}, we consider the case where point 1 is labeled
`in', and assume the value of $\Psi_{\rm R}$ at points 2--8 has been obtained in previous steps
(possibly through the adjustment $\Psi^m\to \Psi^m_{\rm R}=\Psi^m-r\Phi^m_{\rm P}$).
The field $\Psi^m_{\rm R}$ obeys the inhomogeneous equation
$\Box_{\Psi}^m \Psi^m_{\rm R}=Z^m_{\rm R}$, where the source term $Z^m_{\rm R}$
is known analytically, and has a definite finite value everywhere, except on the worldline.
To obtain $\Psi_{\rm R}$ at point 1, we first write finite-difference approximations for
$\Box_{\Psi}^m \Psi^m_{\rm R}$ centered at point 0, as in Eqs.\
(\ref{uv0})--(\ref{average21}). If point 1 is off the worldline, then so is point c, and
we include the source term by just evaluating $Z^m_{\rm R}$ at point c. Solving for
$\Psi^m_{R1}$ yields the finite-difference formula
\begin{equation}\label{fdr2}
\Psi^m_{R1}=[{\rm RHS\ of\ Eq.\ }(\protect\ref{21findif}), {\rm with\ }
\Psi^m_n\to \Psi^m_{Rn}]+h^2 Z^m_{\rm Rc} \quad \text{(in $\cal T$, off the worldline)},
\end{equation}
where $Z^m_{\rm Rc}$ is the value of $Z^m_{\rm R}$ at point c.
This scheme has local discretization error of $\mathcal{O}(h^4)$, with a global
(accumulated) error of $\mathcal{O}(h^2)$, just like the vacuum scheme.

The source $S^m_{\rm R}$, and so also $Z^m_{\rm R}\equiv -(fr/4)S^m_{\rm R}$,
diverges at the worldline in a manner described by the asymptotic formula
(\ref{SRmLocal}), i.e., like $\sim \rho_*^{-1}$, with amplitude depending on
the direction of approach in the $r$--$\theta$ plane. The divergence of the
source poses a technical problem when it comes to numerical implementation:
Even if we re-arrange the grid such that grid points are always avoided by the
worldline, still the rapid growth of the source near the worldline would be
difficult to accommodate numerically. (The problem will show up more acutely
for non-circular, non-equatorial orbits, where it will be more difficult to assure
that the worldline does not ``come too close'' to any of the grid points.)
A natural solution to this problem could be achieved within a higher-order
puncture scheme, in which the puncture function is taken to account for
additional, subdominant terms of the local field. We leave the formulation of
such advanced scheme for future work; here we will continue to use our leading-order
puncture, and demonstrate that even this simple scheme can yield numerically-robust
solutions.

Rather than trying to ``avoid the worldline'' with a suitable layout of grid
points (a strategy which will not be useful anyway for more complicated
orbits), we take here the worldline to cross straight through grid points.
To derive the finite-difference scheme for points crossed by the particle, we
integrate the field equation locally ``by hand'', as we describe in what follows.
This guarantees that $Z^m_{\rm R}$ need never be evaluated at a distance
smaller than $\Delta r_*=h/2$ from the particle. We envisage applying a
similar local-integration procedure for generic orbits, which would save
the need to carefully lay out a ``particle-avoiding'' grid.

\subsection{Treatment of worldline points} \label{wlpts}

Referring, once again, to the grid cell illustrated in Fig.\ \ref{numericalcell}, we
consider the case where point 1 (and so also points c and 4) lie on the
worldline. We assume that the value of $\Psi_{\rm R}^m$ at points 2--8
has been obtained in previous steps, and we need to approximate the value
at point 1. Inside the cell, the source term $Z^m_{\rm R}$ diverges as
$\propto \rho_*^{-1}$. The integaral of $Z^m_{\rm R}$ over the volume of the
cell should therefore yield a finite value. Moreover, we observe in Eq.\
(\ref{SRmLocal}) that the direction-dependence of the leading order,
$\propto \rho_*^{-1}$ divergent term of $Z^m_{\rm R}$ is such that the contribution
from this term vanishes upon integrating over all directions. This suggests a
method for deriving a finite-difference formula for points on the worldline:
Based on the values of  $\Psi_{\rm R}^m$ at points 1--8 of the grid cell, write a
finite-difference approximation for the integral equality
\begin{equation}\label{intgrid}
\int_{\rm cell}\left(\Box_{\Psi}^m \Psi^m_{\rm R}\right) dV=
\int_{\rm cell}Z^m_{\rm R}\mathrm{d}V,
\end{equation}
where the integral is evaluated over the 3-D volume of the grid cell shown in
Fig.\ \ref{numericalcell}, and $\mathrm{d}V= \, \mathrm{d} u\,\mathrm{d}
v\,\mathrm{d}\theta$ is a coordinate (not proper)
volume element; Then solve the resulting discrete algebraic equation for the
value of $\Psi_{\rm R}^m$ at point 1.

To formally discretize the LHS of Eq.\ (\ref{intgrid}), we write
\begin{eqnarray}\label{fff}
\int_{\rm cell}\left(\Box_{\Psi}^m \Psi^m_{\rm R}\right) dV&=&
2\Delta\left(\Psi^m_{R1}-\Psi^m_{R2}-\Psi^m_{R3}+\Psi^m_{R4}\right)
                                                 \nonumber\\
&&-\frac{f_{0}}{8r_{0}^2}\left[
(h^2/\Delta)\left(\Psi^m_{R5} + \Psi^m_{R6} + \Psi^m_{R7}
+ \Psi^m_{R8} - 2\Psi^m_{R2} -2\Psi^m_{R3}\right)  \right.   \nonumber\\
&& \left. -h^2 \Delta \left(2M/r_{0}+m^2\right)
\left(\Psi^m_{R2}+\Psi^m_{R3}\right)  \right]
+\mathcal{O}(h^4\ln h).
\end{eqnarray}
Here $\Psi^m_{Rn}$ represents the numerical value of $\Psi^m_{R}$ at point $n$
of the grid cell shown in Fig.\ \ref{numericalcell}. The error term
symbolizes possible terms of the form $\propto h^3\Delta\ln h$, $h^2\Delta^2\ln h$,
$h^3\Delta\ln \Delta$ and $h^2\Delta^2\ln \Delta$.
In estimating this discretization error we must recall that the field $\Psi^m_{\rm R}$
is continuous, but not differentiable---we expect derivatives of $\Psi^m_{\rm R}$
to diverge logarithmically near the worldline.
The various error terms from the above discretization scheme are expected to be
proportional to $h^3\Delta$ (or $h^2\Delta^2$), {\em times derivatives of
$\Psi^m_{\rm R}$ somewhere in the cell}. Since these derivatives diverge
logarithmically for $h,\Delta\to 0$, we must allow for the above logarithmic error
terms.
Note, finally, that we have dropped the contribution from the term
$\propto \cot\theta\Psi_{{\rm R},\theta}$ in $\Box_{\Psi}^m \Psi^m_{\rm R}$:
This term is already at least of $\mathcal{O}(\Delta\ln\Delta)$ near the particle
[since $\cot\theta\sim -(\theta-\pi/2)$ near $\theta=\pi/2$], and so its volume
integral can be absorbed in the error term in Eq.\ (\ref{fff}).

We next turn to evaluate the RHS of Eq.\ (\ref{intgrid}). We use the
asymptotic
expansion (\ref{SRmLocal}), recalling $Z^m_{\rm R}\equiv -(fr/4)S^m_{\rm
R}$.
The expansion terms specified in Eq.\ (\ref{intgrid}) are sufficient for
our purpose,
since the volume integral of the $\mathcal{O}(\rho_*\ln\tilde\rho_*)$
error term contributes
only at order $\mathcal{O}(h^4\ln h)$ (and higher), which is the order
of error
allowed for on the LHS of Eq.\ (\ref{intgrid}). In terms of the local
polar coordinates
$\rho_*,\phi_*$ defined in Eq.\ (\ref{rhostar}), the volume element is
$\mathrm{d}V=2 f_0^{-1/2}r_0^{-1}\rho_*\mathrm{d}\rho_* \,
\mathrm{d}t\,\mathrm{d}\phi_*$,
and the volume integral reads
\begin{eqnarray}\label{sint1}
\int_{\rm cell}Z^m_{\rm R}\mathrm{d}V=
-\frac{q}{4\pi r_0 f_0^{1/2}}e^{-im\omega t_{\rm c}}
\int_{\rm cell}
\mathrm{d}\rho_* \, \mathrm{d}t\,\mathrm{d}\phi_* rf
\left[\alpha(\phi_*)+\beta_{\rm ln}\rho_*\ln(\tilde\rho_*/4)+
\rho_*\beta^m(\phi_*)\right] +\mathcal{O}(h^4\ln h),
\end{eqnarray}
where $t_{\rm c}$ is the value of $t$ at the central point c. Here we
applied
a ``constant phase'' approximation, which is valid since, for $m$ not
too large,
the factor $e^{-im\omega t}$ varies negligibly across one grid cell.
We next expand the factor $rf=r-2M$ in the integrand as
$rf=r_0f_0+\delta r\cong r_0f_0 + f_0\delta r_* = r_0f_0 + f_0^{1/2} \rho_* \cos\phi_*$, with higher-order terms of
the expansion
absorbed in the error term $\mathcal{O}(h^4\ln h)$ of the integral.
We observe, recalling $\alpha(\phi_*)\propto\cos\phi_*\sin^2\phi_*$, that
the term $\propto(r_0f_0)\alpha(\phi_*)$ vanishes upon integrating
$\int_{0}^{2\pi}\mathrm{d}\phi_*$. All $\phi_*$-dependent terms in
$(r_0f_0)\beta^m(\phi_*)$ similarly yield a vanishing contribution.
Omitting higher-order terms, we are left with
\begin{eqnarray}\label{I1}
\int_{\rm cell}Z^m_{\rm R}\mathrm{d}V=
-\frac{q}{4\pi r_0}e^{-im\omega t_{\rm c}}
\int_{\rm cell}
\mathrm{d}\rho_* \, \mathrm{d}t\,\mathrm{d}\phi_* \rho_*
\left[\alpha(\phi_*)\cos\phi_*
+r_0f_0^{1/2}\left(\beta_{\rm ln}\ln(\tilde\rho_*/4)+
\beta_0^m\right)\right].
\end{eqnarray}
This integral can be evaluated explicitly---most easily by first
transforming
back to local Cartesian coordinates $\delta r_*,\delta\theta$. We obtain
\begin{eqnarray}\label{I2}
\int_{\rm cell}Z^m_{\rm R}\mathrm{d}V &=&
\frac{q\Delta h^2}{24\pi}r_0f_0e^{-im\omega t_{\rm c}}
\left\{\alpha_0+11\beta_{\rm ln}-6\beta_0^m-(\alpha_0+2\beta_{\rm ln})
\left[3\zeta\arctan(1/\zeta)+(1/\zeta)\arctan\zeta\right]
\right. \nonumber\\
&&+\left.
2\left[2\zeta^2\alpha_0+\beta_{\rm ln}(\zeta^2-3)\right]
\ln\left(\zeta^{-1}\sqrt{1+\zeta^2}\right)
-6\beta_{\rm ln}\ln\left(r_0\Delta/(8P_{\varphi\varphi}^{1/2})\right)
\right\}\nonumber\\
&\equiv&2\Delta h^2 \tilde Z^m_{\rm Rc}(t_{\rm c}),
\end{eqnarray}
where
\begin{equation}
\alpha_0=\frac{8(1-M/r_0)}{r_0^3 f_0^{1/2} {\cal E}}, \quad \quad
\zeta=2r_0f_0^{-1/2}(\Delta/h).
\end{equation}

Finally, equating Eqs.\ (\ref{fff}) and (\ref{I2}) and solving for $\Psi_{\rm R1}^m$,
we obtain the finite-difference formula for worldline points:
\begin{equation}\label{fds3}
\Psi^m_{R1}=[{\rm RHS\ of\ Eq.\ }(\protect\ref{21findif}), {\rm with\ }
\Psi^m_n\to \Psi^m_{Rn}]+h^2 \tilde Z^m_{\rm Rc} \quad \text{(on the worldline)},
\end{equation}
with a local error of $\mathcal{O}(h^3\ln h)$.

We can summarize our finite-difference scheme for any grid point as follows.
Given the values of the numerical field at points 2--8 of the grid cell depicted
in Fig.\ \ref{numericalcell}, and assuming these values have already been adjusted
as either all `in' or all `out' (as described above), then the value of the field
at point 1 is approximated by
\begin{equation}
\Psi^m_{1}=[{\rm RHS\ of\ Eq.\ }(\protect\ref{21findif})]+
\left\{\begin{array}{lll}
0 &+ \mathcal{O}(h^4),                      & \text {point 1 outside $\cal T$,} \\
h^2 Z^m_{\rm Rc}&+\mathcal{O}(h^4),       & \text {point 1 inside $\cal T$, off worldline,} \\
h^2 \tilde Z^m_{\rm Rc}&+\mathcal{O}(h^3\ln h), & \text {point 1 inside $\cal T$, on worldline}.
\end{array}\right.
\end{equation}
The global (accumulated) error from points off the worldline is expected to be
$\mathcal{O}(h^2)$. The error from worldline points accumulates only along the worldline
(the number of points contributing to it scales as $\sim 1/h$), and is expected
to dominate the global evolution error, with contribution of $\mathcal{O}(h^2\ln h)$.
{\it We thus expect our scheme to converge at least linearly, but note that the above
logarithmic error terms are likely to deter quadratic convergence.}

How could one eliminate the dominant logarithmic error terms, in order to assure
quadratic convergence? The occurrence of such terms can be traced back to the
logarithmic divergence of the R-field derivatives at the worldline, which is a feature
of the leading-order puncture scheme adopted here. A natural solution to the problem could be
offered within a higher-order puncture scheme, which incorporates a differentiable
($C^1$) R-field. We leave the formulation of such a scheme for future work.

\subsection{Boundary and initial conditions}

At the poles ($\theta=0,\pi$) we apply boundary conditions as in the vacuum case---see
Eq.\ (\ref{BC}). For initial conditions, we simply set the field (both $\Psi$ and
$\Psi_{\rm R}$) to zero along $u=u_0$ and $v=v_0$. This produces a burst of spurious
radiation, mainly at the particle's initial location and at the intersection of
$\partial \cal T$ with the initial surfaces. The spurious waves die off at late time,
gradually unveiling the physically-meaningful field. In application of the code,
one must monitor the effect of residual spurious waves, by testing the stationarity
of the late-time field.

\section{Results and Code validation}\label{Sec:VI}

\subsection{Sample results}
\begin{figure}
\includegraphics[width=0.49\textwidth]{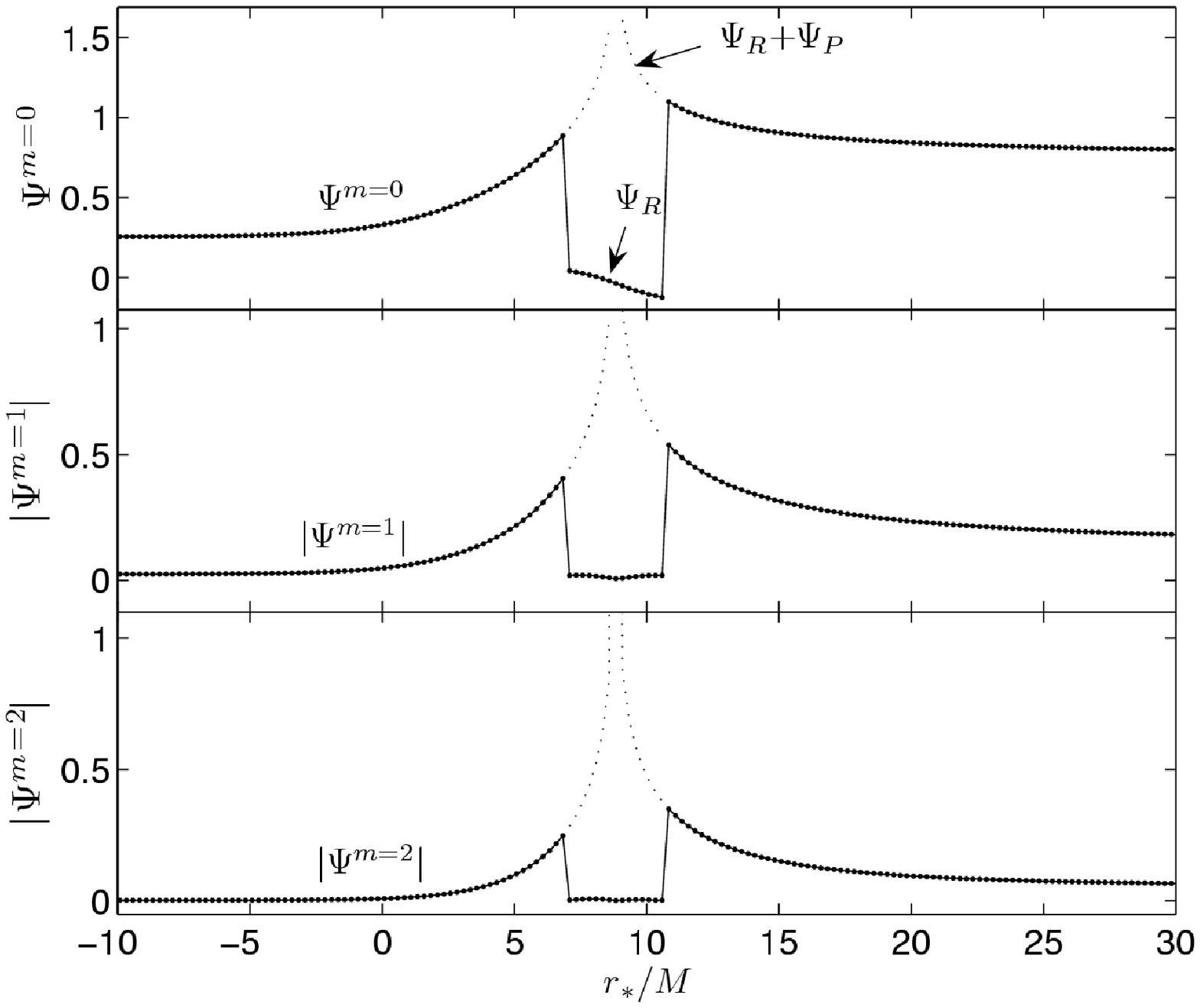}
\includegraphics[width=0.49\textwidth]{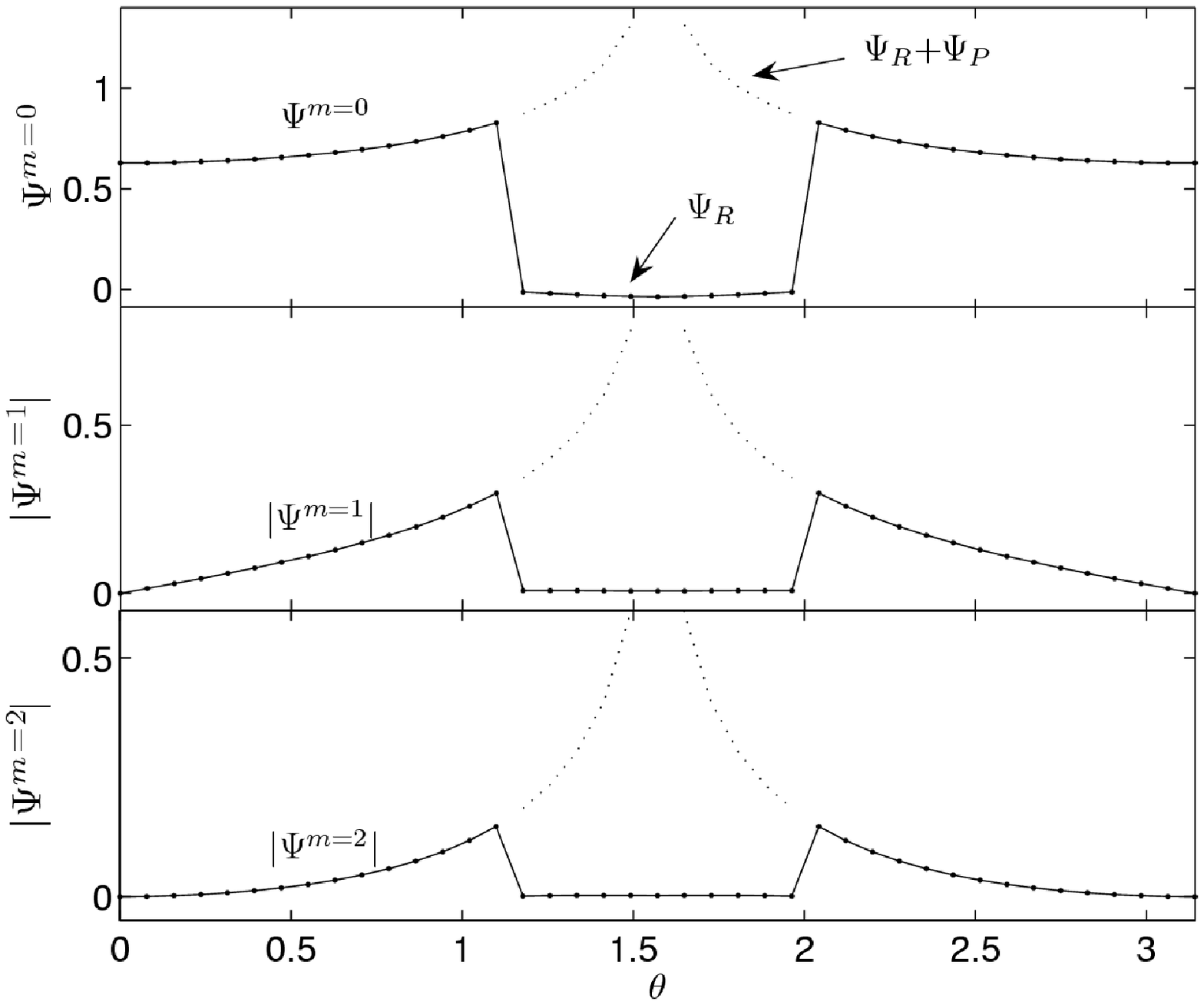}
\includegraphics[width=0.49\textwidth]{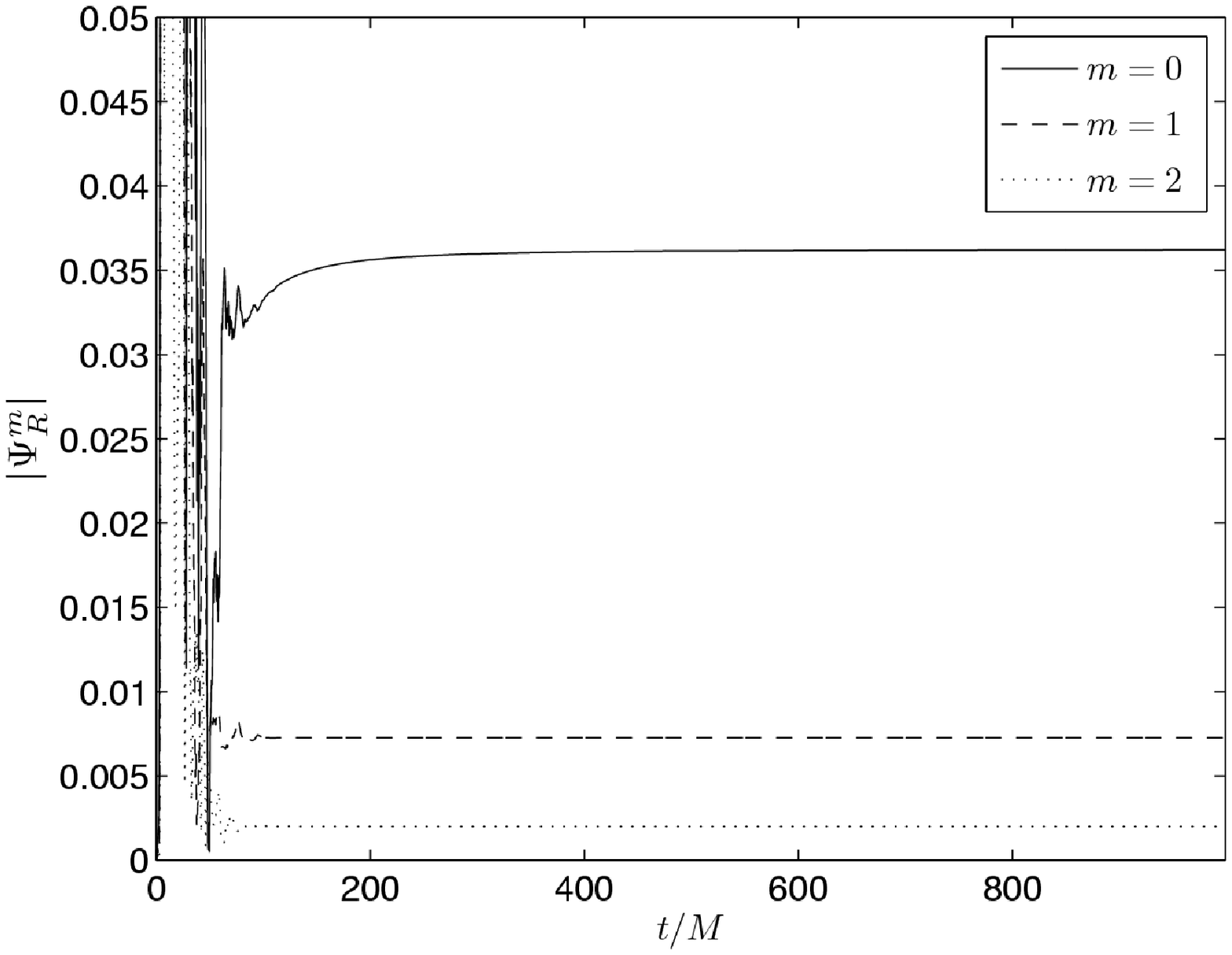}
\caption{\label{fig:solutions}
Sample numerical results for $r_0=7M$ ($r_{*0}\cong 8.8326M$), showing each of the 3
modes $\Psi^{m=0,1,2}$ along 3 different slice-cuts of the 2+1-D domain:
The upper left figure shows $|\Psi^m(r_*)|$ at $\theta=\pi/2$ and $t=500M$;
the upper right figure shows $|\Psi^m(\theta)|$ at $r=7M$ and $t=500M$; and
the lower figure shows $|\Psi_{\rm R}^m(t)|$ at $r=7M$ and $\theta=\pi/2$
(namely, along the particle's worldline). The dimensions of the auxiliary worldtube here are
$\delta_{r_*}=7.5M$ and $\delta_{\theta}=\pi/4$. In the two spatial slices we display
the full field $\Psi^m$ outside the worldtube, and the residual field $\Psi^m_{\rm R}\equiv\Psi^m
-\Psi^m_{\rm P}$ inside it. (Recall $\Psi^m_{\rm P}$ is the puncture function, given
analytically.) Inside the worldtube we also indicate, in dotted line,
the full (divergent) field, obtained through $\Psi^m=\Psi^m_{\rm R}+\Psi^m_{\rm P}$.
In the two spatial slices, dots along the graphs mark the location of
actual numerical grid points. The various features of these solutions are discussed in the text.
}
\end{figure}

We present results for several sample cases, highlighting a few generic features of
the numerical solutions generated using the above puncture scheme. The plots in
Fig.\ \ref{fig:solutions} show numerical results for the modes $m=0,1,2$ of the scalar
field $\Psi^m$, for a circular geodesic orbit with radius $r_0=7M$ ($r_{*0}\cong 8.8326M$).
All solutions were obtained with a grid resolution of $h=M/4$ and $\Delta=\pi/40$,
giving a ratio $\Delta/h\sim 0.31M^{-1}$---just above the Courant limit.
(Unlike in the vacuum case, near the particle the resolution requirement in the
longitudunal direction is as high as in the radial direction, which requires us to
lower the ratio $\Delta/h$.) The worldtube dimensions were taken as
$\delta_{r_*}=7.5M$ and $\delta_{\theta}=\pi/4$. The evolution starts at $t=0$ and ends
at $t=1000M$ (covering roughly 8 orbital periods).

We highlight a few of the features visible in these plots:
(i) The early stage of the numerical evolution is dominated by noise from
spurious initial waves. These die off within $\sim 1$ orbital period, giving way
to stationary behavior at late time.
(ii) The full field calculated outside the worldtube merges smoothly with
$\Psi^m_{\rm R}+\Psi^m_{\rm P}$ across the boundaries of the worldline, as expected.
(iii) The numerical variable $\Psi^m_{\rm R}$ is continuous at the particle, and
has a well-defined value there. This, of course, makes it much more tractable
numerically than the original, divergent field $\Psi^m$.
(iv) The value of $\Psi^m_{\rm R}$ at the particle drops rapidly with increasing $m$.
(v) The residual field $\Psi^m_{\rm R}$ is asymmetric about the particle in the
radial direction, reflecting the slight anisotropy of the curved background
spacetime. It is this asymmetry in the field that gives rise to the SF effect.
A scheme for constructing the physical SF from $\Psi^m_{\rm R}$ will be
presented elsewhere.

\subsection{Tests of code}

In what follows we demonstrate the numerical robustness of our code by
(i) demonstrating point-wise numerical convergence, and (ii) showing that our
numerical solutions depend only weakly on the dimensions of the auxiliary worldtube,
and that this dependence gets ever weaker with improving resolution.
We then validate our numerical solutions (and the entire puncture scheme) by
comparing with results obtained from our 1+1-D companion code.

\subsubsection{Numerical convergence}

\begin{figure}
\includegraphics[width=0.49\textwidth]{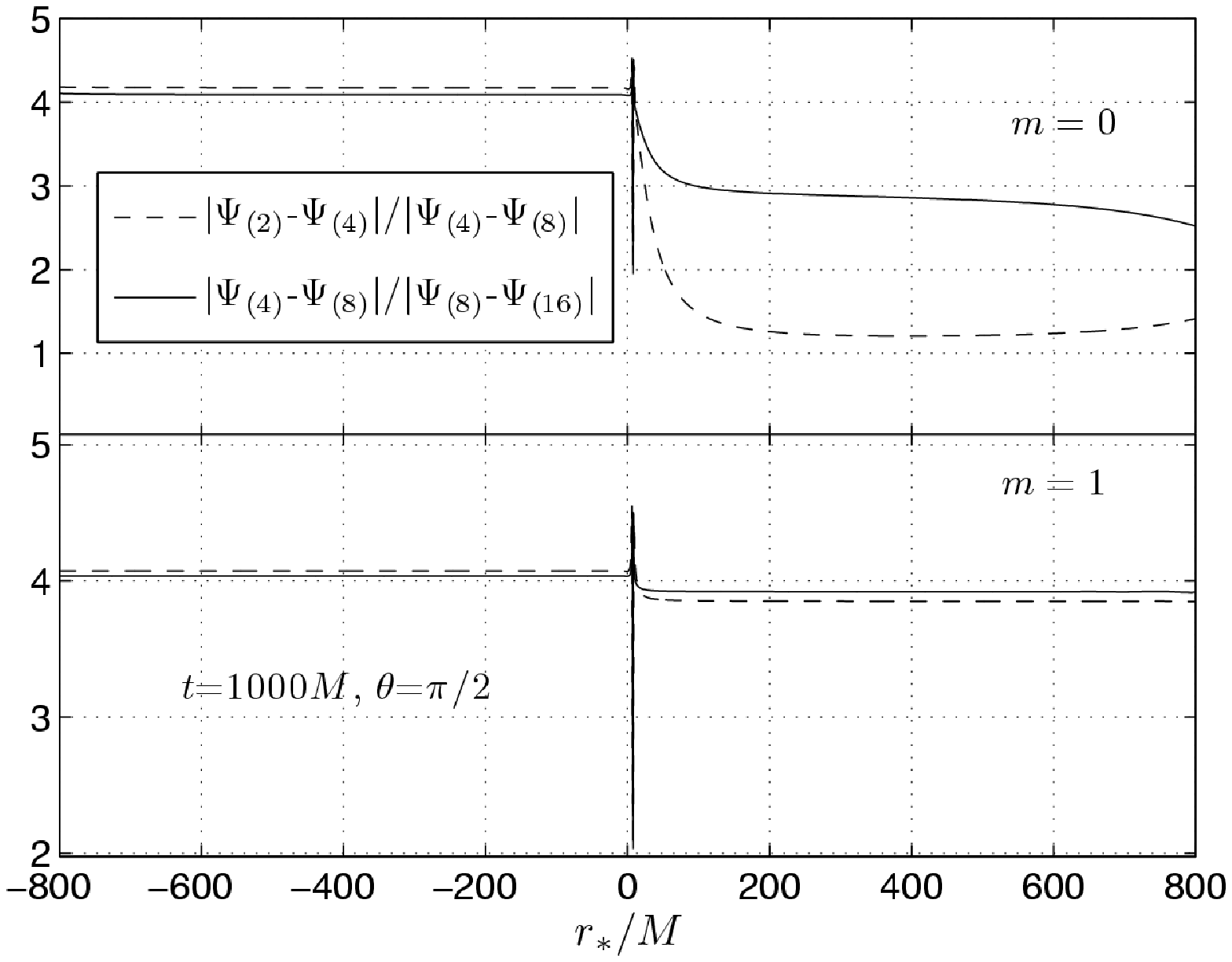}
\includegraphics[width=0.49\textwidth]{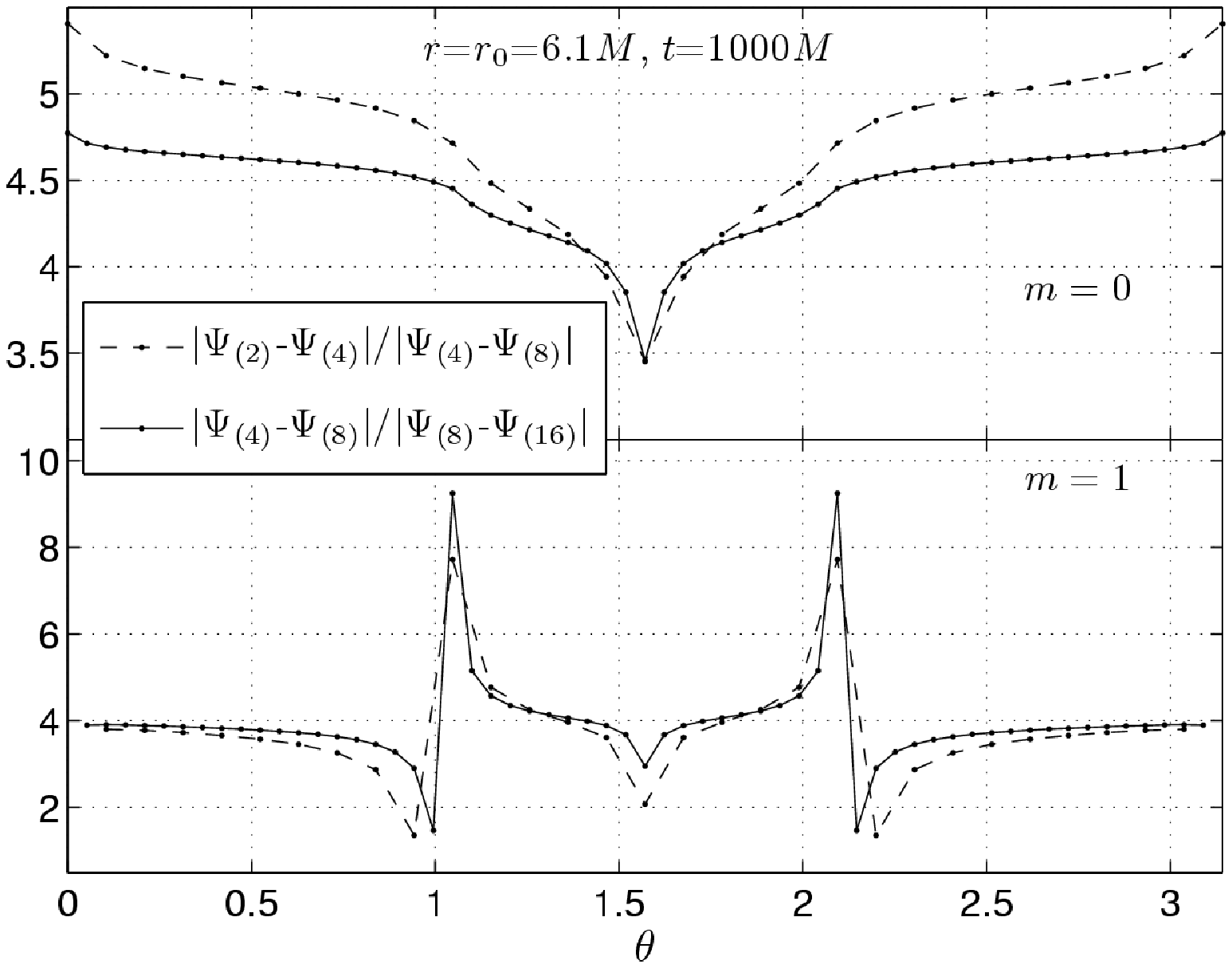}
\includegraphics[width=0.49\textwidth]{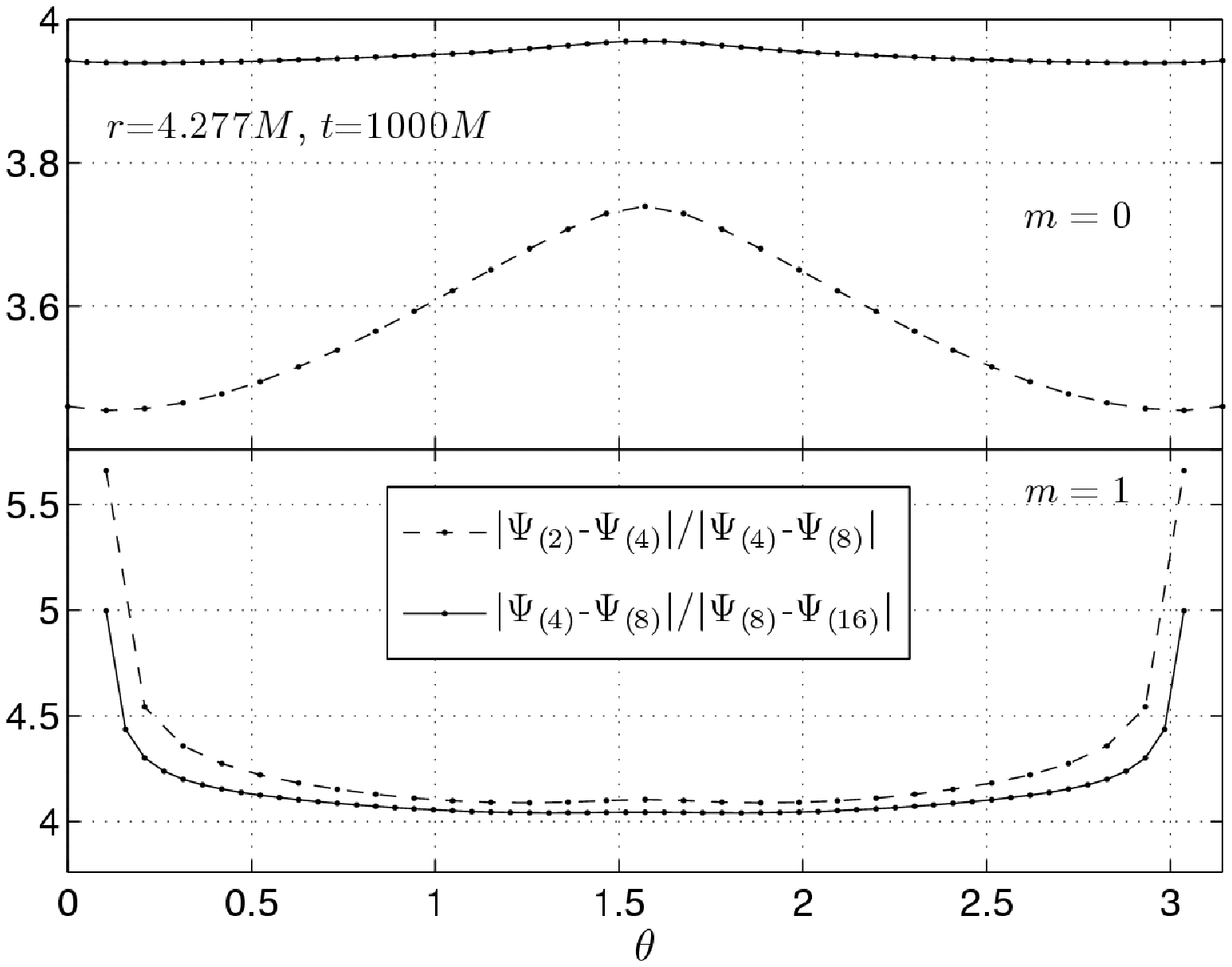}
\includegraphics[width=0.49\textwidth]{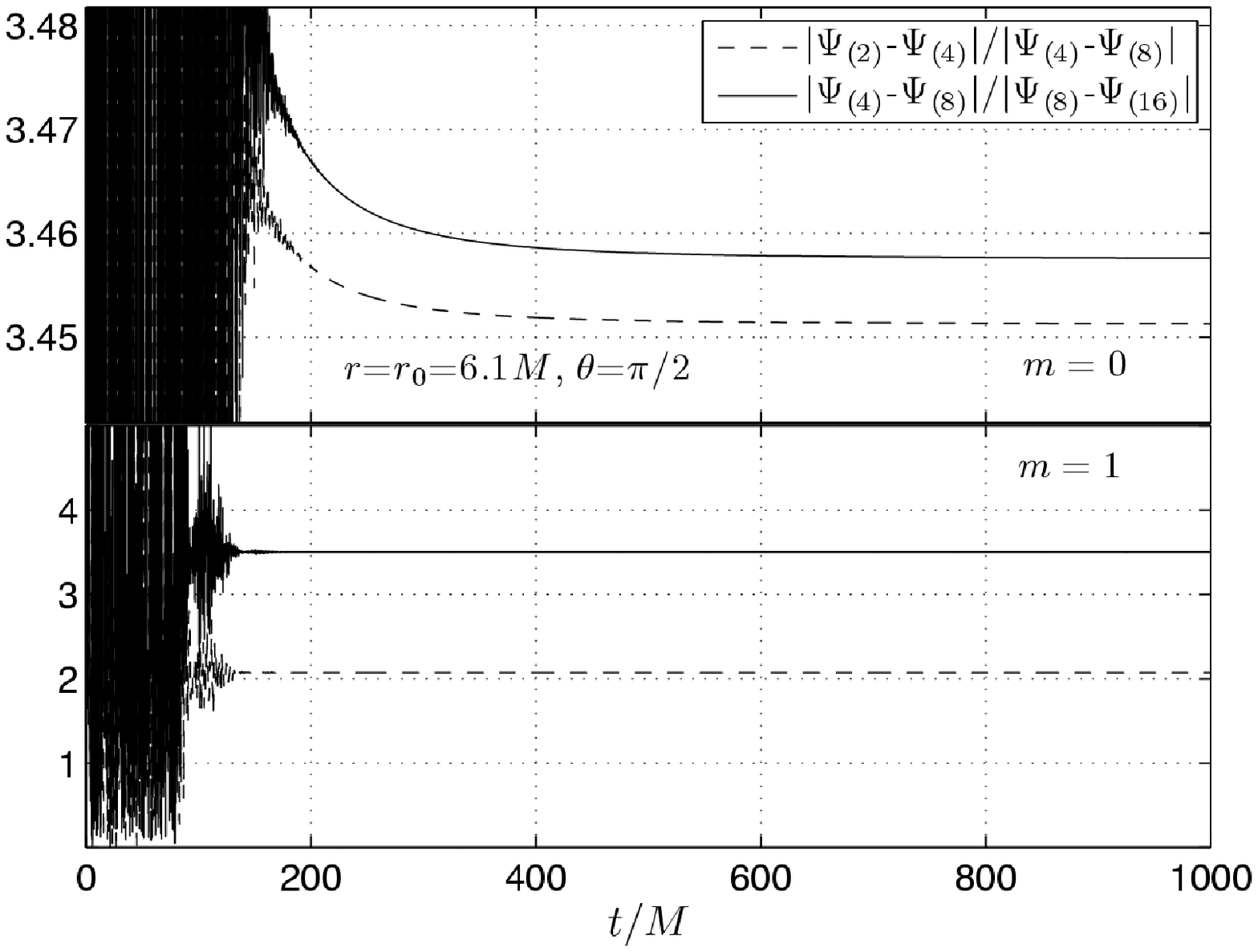}
\caption{\label{fig:conv2} Numerical convergence test for the puncture scheme.
Here we set $r_0=6.1M$ ($r_{*0}\cong 7.5357M$), and take worldtube dimensions of
$\delta_{r_*}=5M$, $\delta_{\theta}=\pi/3$.
We plot the relative differences $\delta\Psi_{\rm rel}^m\equiv\left
|(\Psi^m_{(n)}-\Psi^m_{(2n)})/(\Psi^m_{(2n)}-\Psi^m_{(4n)})\right|$, for
$n=2,4$, where $\Psi^m_{(n)}$ is the solution obtained with resolution
$h=(10M/\pi)\Delta=M/n$. Outside the worldtube we show the relative differences
in the full field $\Psi^m$, and inside the worldtube---in the regular
variable $\Psi_{\rm R}^m$. {\it Upper-left panel:}
$\delta\Psi_{\rm rel}^m(r_*)$ for $(t,\theta)=(1000M,\pi/2)$ (crossing the
particle).
{\it Upper-right panel:} $\delta\Psi_{\rm rel}^m(\theta)$ for
$(t,r)=(1000M,r_{0})$ (crossing the particle).
{\it -left panel:} $\delta\Psi_{\rm rel}^m(\theta)$ for
$(t,r)=(1000M,4.2766M)$ (off the particle).
{\it Lower-right panel:}
$\delta\Psi_{\rm rel}^m(t)$ for $(\theta,r)=(\pi/2,r_{0})$, i.e., along
the particle's worldline.
Away from the particle, the convergence is approximately quadratic;
near the particle the convergence rate is harder to characterize, but remains
at least linear.
}
\end{figure}

As in the vacuum case, we examined the point-wise convergence of our scheme
by comparing solutions obtained with different grid resolutions.
Figure \ref{fig:conv2} demonstrates the convergence properties of our solutions,
for $r_0=6.1M$ and $m=0,1$. The following features are manifest:
(i) Away from the particle, and after the decay of the initial spurious
waves, the numerical solutions show an approximate quadratic convergence.
(ii) Near the particle (within a distance of a few $M$) the convergence
is not uniform, and more difficult to characterize---but appears to be
better than linear everywhere.
(iii) On the particle itself, the convergence seems, once again, slower
than quadratic and faster than linear.
We have confirmed the generality of these features by experimenting
with a range of orbital radii $r_0$ and modes $m$, and different
auxiliary worldtube dimensions.
We suspect that what slows down the convergence near the
particle are the logarithmic error terms discussed above. Possible ways to improve the
convergence of the scheme in future work will be discussed below.

\subsubsection{Dependence on worldtube dimensions}

\begin{figure}
  \includegraphics[width=0.49\textwidth]{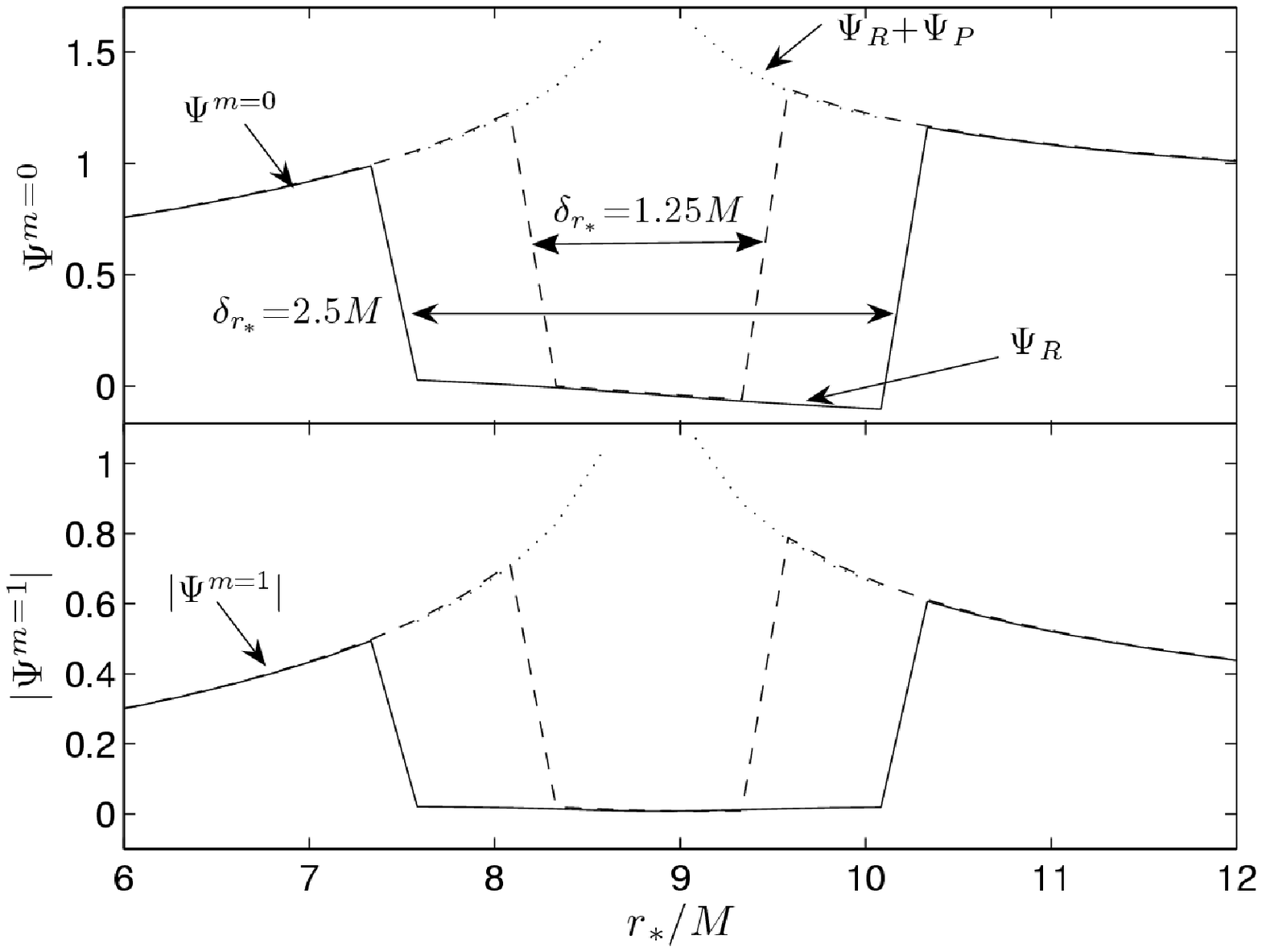}
  \includegraphics[width=0.49\textwidth]{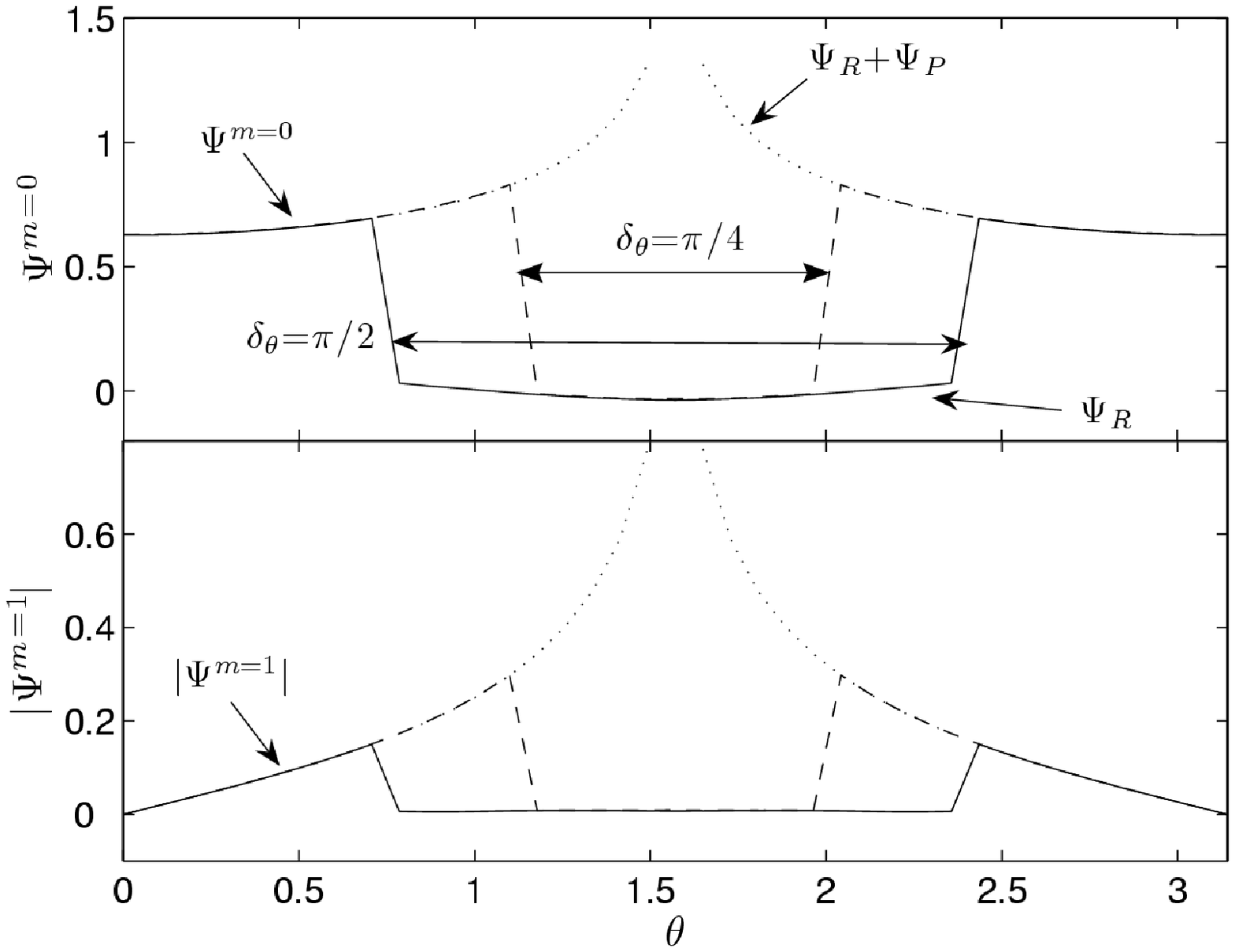}
\caption{\label{fig:wtcompm0}
Independence of the numerical solutions on the choice of worldtube
dimensions---demonstrated here for $r_0=7M$, $m=0,1$. The graphs
compare the solutions obtained using two different auxiliary worldtubes:
one with dimensions $(\delta_{r_*},\delta_{\theta})=(1.25M,\pi/4)$ (dashed line),
and the other with dimensions $(\delta_{r_*},\delta_{\theta})=(2.5M,\pi/2)$
(solid line). The dotted line is the full solution $\Psi^m_{\rm R}+\Psi^m_{\rm P}$,
as obtained with the larger worldtube. The left panel displays the behavior as a
function of $r_*$ at $(t,\theta)=(500M,\pi/2)$, and the right panel shows the
behavior as a function of $\theta$ at $(t,r)=(500M,r_0)$. The two calculations
agree well on the value of $\Psi_{\rm R}$ inside the small worldtube, and on
the value of $\Psi^m$ elsewhere. It is demonstrated in Fig.\ \ref{fig:wtconv}
below (for the $m=0$ case) that the tiny discrepancy between the two solutions
tends to zero with increasing grid resolution.
}
\end{figure}

\begin{figure}
  \includegraphics[width=0.49\textwidth]{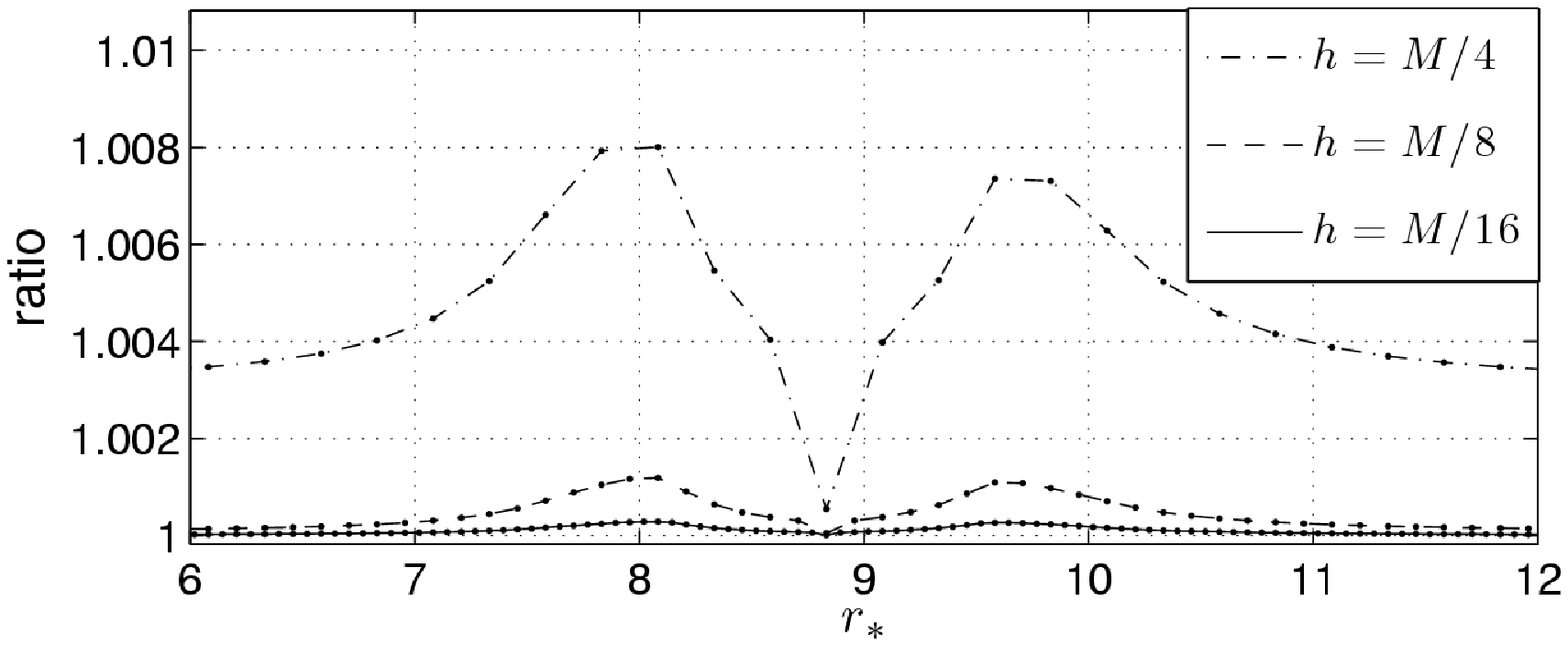}
  \includegraphics[width=0.49\textwidth]{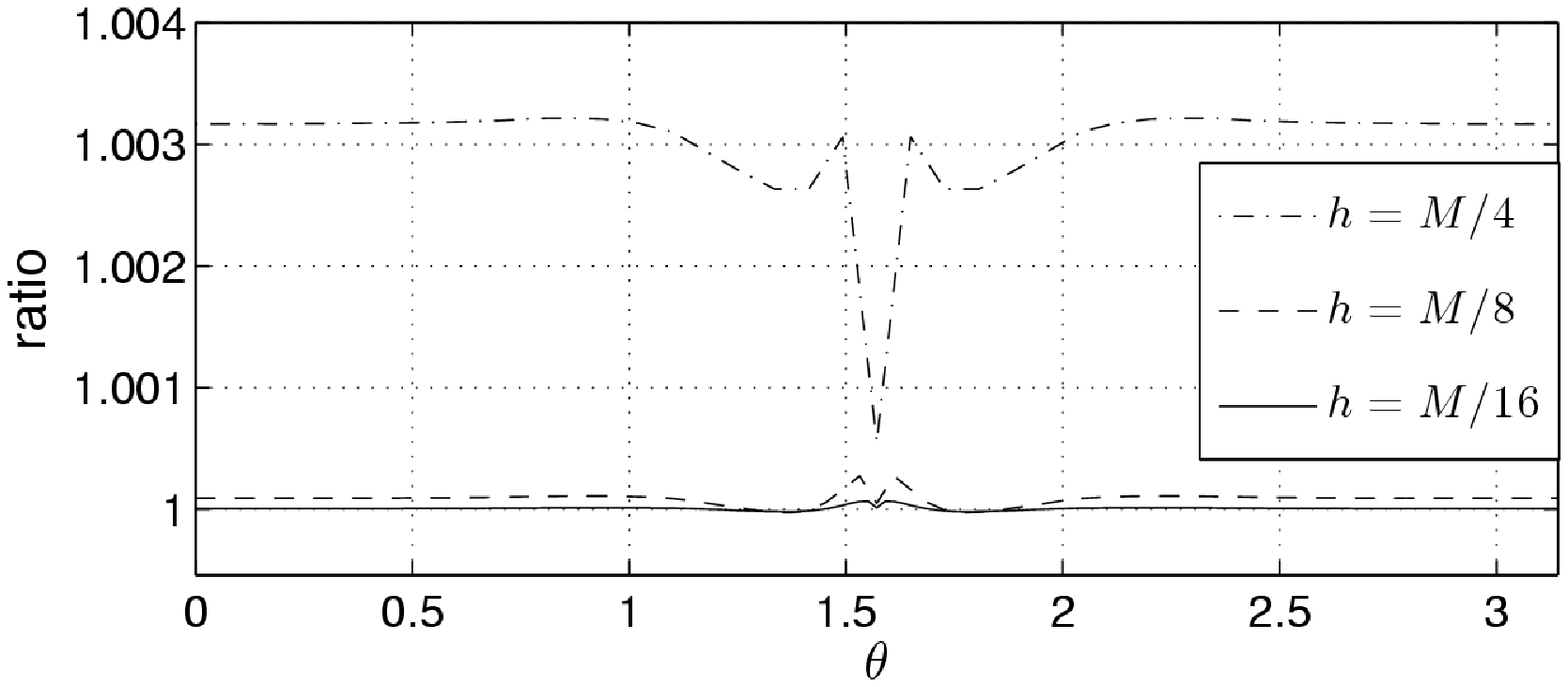}
\caption{\label{fig:wtconv}
Data corresponding to the two upper plots
of Fig.\ \ref{fig:wtcompm0}: Shown here, for $m=0$, is the {\em ratio} between
the two numerical solutions obtained with different worldtubes (i.e., the
ratio between the dash and solid lines in Fig.\ \ref{fig:wtcompm0}).
The ratio is displayed for three different grid resolutions [$h=M/4,M/8,M/16$,
where in each case $\Delta=(\pi/10)M^{-1}h$]. The small discrepancy between the
two solutions diminishes rapidly with increasing resolution, suggesting that
our numerical solutions are insensitive to the choice of auxiliary worldtube
at the continuum limit, as should be expected.}
\end{figure}

It is important to establish that our numerical solutions do not depend on
the dimensions of the auxiliary wolrdtube (modulo discretization error).
We have tested the code with various worldtube dimensions, and present
typical results in Figs.\ \ref{fig:wtcompm0} and \ref{fig:wtconv}. The figures
compare between solutions obtained using two different worldtubes: One
with dimensions $(\delta_{r_*},\delta_{\theta})=(1.25M,\pi/4)$, and the
other with dimensions $(\delta_{r_*},\delta_{\theta})=(2.5M,\pi/2)$.
Evidently, the value of the calculated field is only very slightly affected
by the different choice of worldtube, and the tiny differences appear to
diminish rapidly with improving resolution. Figs.\ \ref{fig:wtcompm0} and
\ref{fig:wtconv}
demonstrate this behavior for $r_0=7M$ and $m=0,1$, but we observe similar
behavior for other $r_0$ and $m$.


\subsubsection{Comparison with 1+1-D solutions}

\begin{figure}
\includegraphics[width=0.49\textwidth]{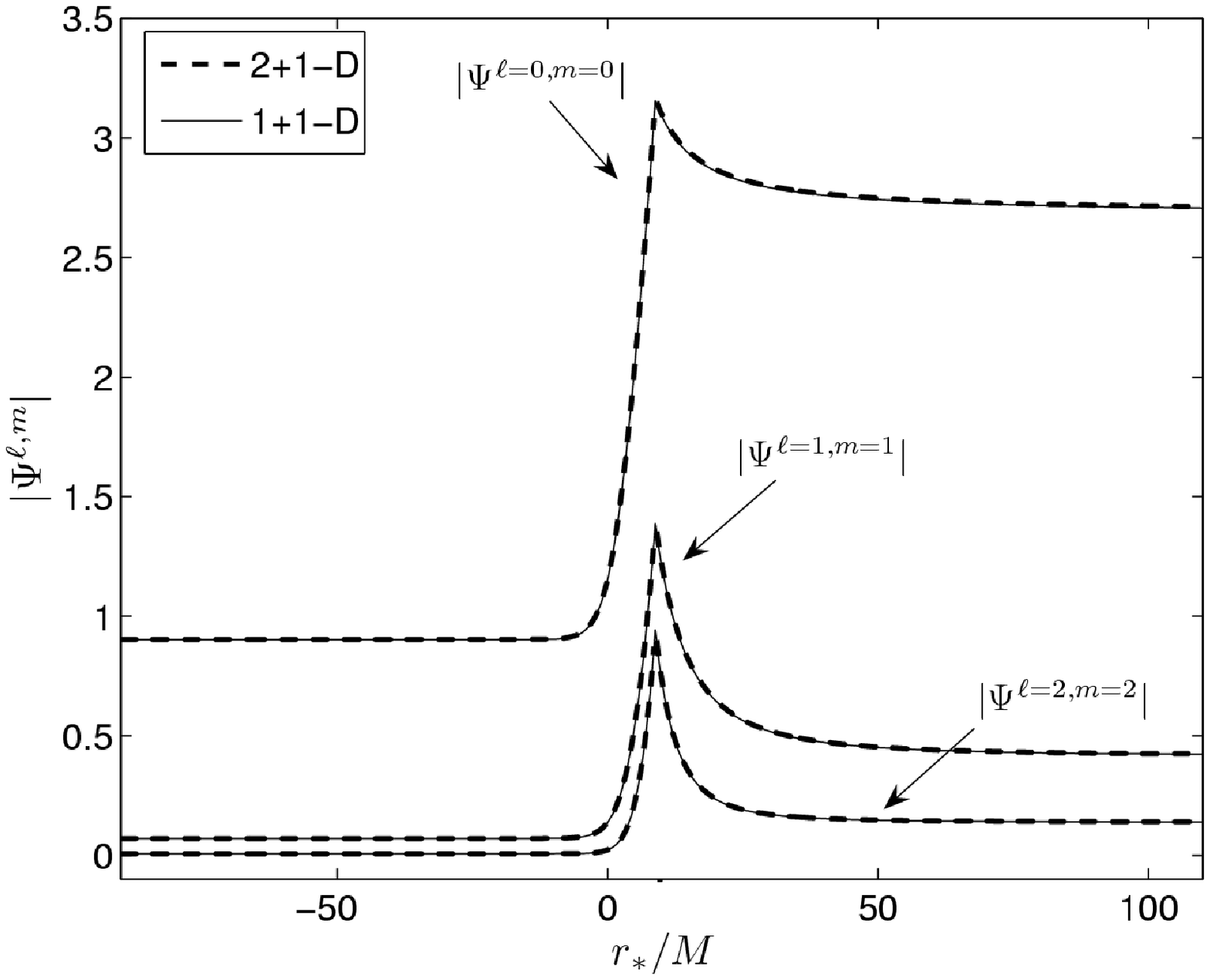}
\includegraphics[width=0.49\textwidth]{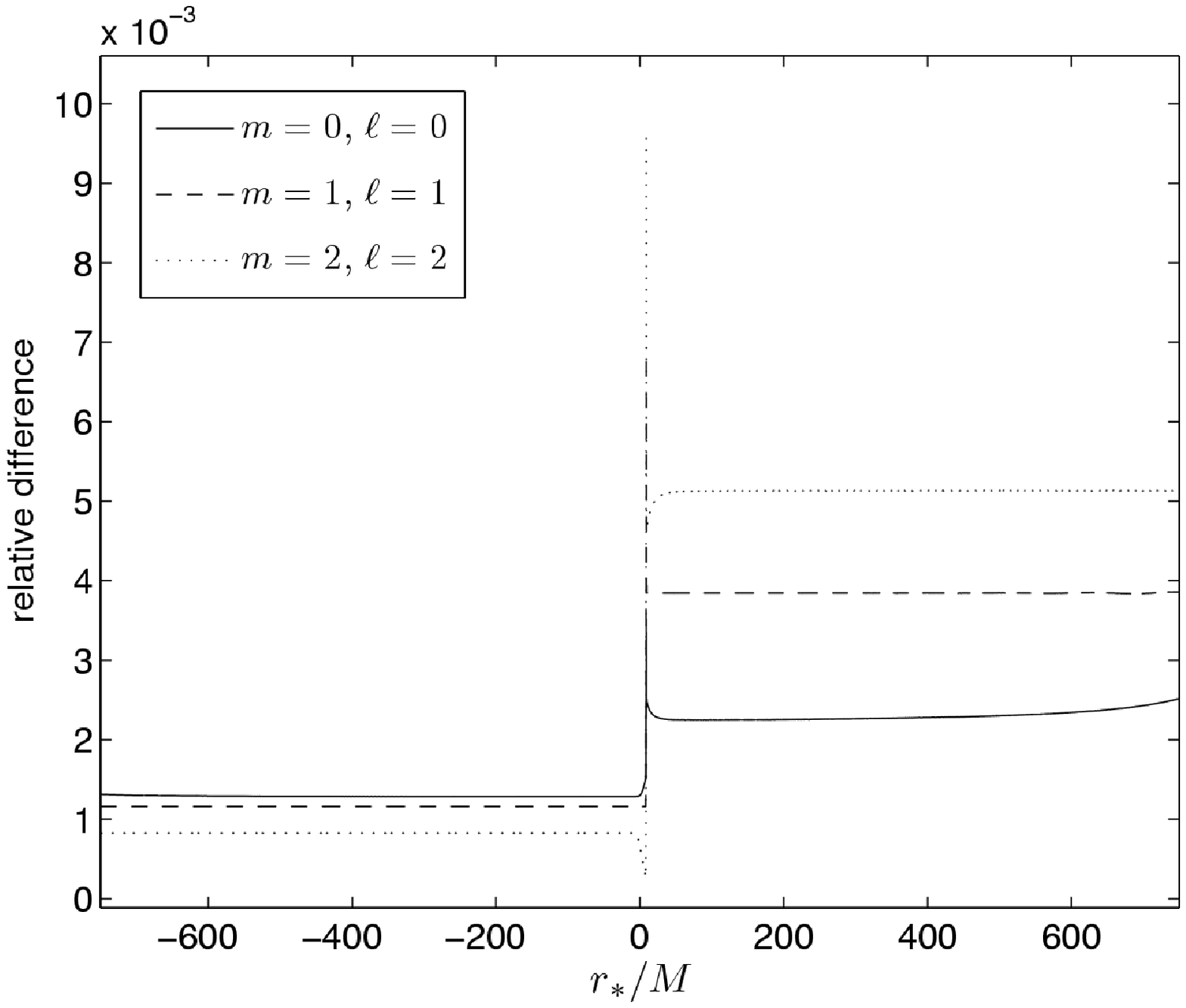}
\caption{\label{fig:sourcecomp1112}
Left panel: Comparison of 2+1-D solutions (dashed line) with 1+1-D solutions
(solid line), for individual multipole modes $(\ell,m)=(0,0),(1,1),(2,2)$.
The 2+1-D modes were obtained by decomposing the original 2+1-D solutions
into their individual $\ell$ components (using numerical integration).
The data here is for $r_0=7M$, and the field is extracted at $t=1000M$ in
both codes.
Right panel: The relative difference between the 2+1-D and 1+1-D solutions
shown on the left. All modes compare to within less than 1\% difference
(or far better), providing a strong validation test for the 2+1-D
code.}
\end{figure}

A good quantitative test of our code is provided by comparing the
2+1-D solutions with solutions obtained using a 1+1-D evolution code.
To obtain 1+1-D solutions for a scalar charge in a circular orbits,
we extended our vacuum 1+1-D code to incorporate a source particle,
using the prescription of Ref.\ \cite{BB2000}. The code calculates
individual multipole modes $\ell,m$ of the scalar field. To allow comparison
with the 2+1-D code, we decomposed the 2+1-D numerical solutions into
their individual $\ell$ modes (by integrating numerically with respect to
$\theta$ against individual Legendre functions with given $\ell,m$).
Results from such comparison, for $r_0=7M$ and $m=0,1,2$, are shown
in Fig.\ \ref{fig:sourcecomp1112}. In all cases examined we find convincing
agreement between the 2+1-D and 1+1-D solutions.

\section{Summary and discussion}\label{Sec:VII}

In this work we began developing the computational framework to facilitate
evolution of black hole perturbations from point particles in 2+1-D. This
is mainly motivated within the context of SF calculations in Kerr:
Knowledge of the perturbation field near the particle is an essential input
for any calculation of the local SF. The local field near a particle
in Kerr orbits has been studied so far mainly in a 1+1-D framework (i.e.,
through a spherical-harmonic decomposition), and so we started our analysis by
exploring the field behavior in 2+1-D. Specializing to a scalar field, we
established that each azimuthal $m$-mode of the perturbation generically shows a
logarithmic divergence near the particle. We then devised a numerical evolution
scheme for the scalar field, based on approximating the divergent piece of
the field analytically, and solving for the finite (and continuous) residual
part. We demonstrated the applicability of this ``puncture'' scheme in the test
case of circular orbits in Schwarzschild (but working in 2+1-D). For this,
we developed a new 2+1-D evolution code, which we tested for numerical
robustness and by demonstrating agreement with solutions obtained using other
methods. We found that the scheme successfully resolves the residual,
sub-dominant behavior of the scalar field near the particle.
Whether our code, in its current form (and with realistic numerical
resolution), allows sufficient accuracy for precise SF calculations
remains to be explored, and we leave this for future work.

The main strength of our time-domain approach is in the fact that it is
now rather straightforward to extend the analysis to problems which are more
astrophysically interesting. Our scalar field code can be readily extended to
deal with inclined and eccentric orbits, and generalization to Kerr spacetime
could be achieved rather straightforwardly based on the existing platform.
We envisage applying the same numerical approach for solving the gravitational
perturbation equations (in the Lorenz gauge), but we anticipate this would
require much preparatory formulation work (to cast the equations in a form
suitable for 2+1-D evolution).

Before attempting further extensions/applications of the code, one may consider
a few possible improvements of the numerical method. Firstly, working with a
{\em higher-order} puncture scheme could prove very beneficial.
In our leading-order scheme, the regularized field $\Psi^m_{\rm R}$ is not
differentiable at the location of the particle (derivatives of $\Psi^m_{\rm R}$
diverge there logarithmically, in a direction-dependent manner), which
complicates the analysis. This can be cured by including higher-order terms in the
definition of the analytic puncture $\Psi_{\rm P}$. A second-order puncture can
readily be constructed based on Eq.\ (\ref{2ndorder}) above, which should
yield a differentiable field $\Psi^m_{\rm R}$ (with nearly no extra cost
in computation time). This should have the following benefits: (i) The source
term in the $R$-field equation would no longer have a strong, $\rho_*^{-1}$
divergence near the particle---it would instead diverge logarithmically, which
is much easier to accommodate numerically. (ii) No logarithmic error terms
of the sort discussed in Sec.\ \ref{wlpts} would occur, which should settle the
quadratic convergence of the scheme near the particle. (iii) SF
calculations require the field's {\em derivatives} at the particle.
The $R$-field calculated using our leading-order puncture would therefore require
further regularization, whereas, at least in principle, the SF should
be accessible directly from a differentiable $R$-filed coming from a
second-order puncture scheme (cf.\ discussion below).

Another possible improvement concerns the choice of initial data for the evolution.
Expediting the damping of the spurious initial waves would allow shorter
runs and save in computational cost. An improved scheme could incorporate
smooth approximate initial data. Another idea, easier to implement, is to use
interpolated solutions from low-resolution runs as initial conditions for
high-resolution evolution. We are planning to incorporate the latter scheme in
future applications of our code.

Finally, we briefly discuss the application of our method to SF calculations.
The standard ``mode-sum'' formula for the SF in Kerr \cite{Barack:1999wf} requires
as input the individual $\ell,m$ multipole modes of the perturbation field.
To apply the mode-sum formula in its standard form, we would therefore have to
decompose our numerical $m$-mode solutions into their individual $\ell$-mode
components. A more direct approach would be to access the SF directly
from our regular $m$-mode fields $\Psi^m_{\rm R}$. We are currently formulating
such an ``$m$-mode sum'' scheme for the SF, and will present it
elsewhere \cite{BGS2007}. The proposed scheme could use either the
first-order puncture solutions $\Psi^m_{\rm R}$ calculated here (which would
then require us to apply a certain local-averaging procedure), or, in its simpler
form, it could use the solutions of a future second-order puncture code.

\section{Acknowledgments}

We thank Carsten Gundlach, Ian Hawke, Norichika Sago, Jonathan Thornburg, and James Vickers
for their helpful input. The basic idea for this project stemmed from
discussions with Richard Price and Napoleon Hernandez, for which we are grateful to both.
Napoleon Hernandez's master thesis (University of Texas at Brownsville, unpublished)
explores the implementation of a similar idea.
This work was supported by PPARC through grant number PP/D001110/1.
We also gratefully acknowledge financial support from the Nuffield Foundation.

\appendix
\section{Tables of polynomials} \label{AppA}

We tabulate here the various polynomials appearing in the expressions for
$\Phi_{\rm P}^m$ [Eq.\ (\ref{PhiPm})] and $S_{\rm R}^m$ [Eq.\ (\ref{SRm2})
with Eq.\ (\ref{Im})], for the 6 modes $m=0$--$5$.
The polynomials $p_{K}^m$ and $p_{E}^m$ are listed in Table \ref{Table:pKpE}.
The polynomials $p_{nK}^m$ and $p_{nE}^m$ (with $n=1$--$4$) are listed in
Table \ref{Table:pKpEn}.

\begin{table}
\centering
\begin{displaymath}
\begin{array}{c|l}
m & p_K^m \\
\hline\hline
0 & 2 \\
1 & 2(1+2\tilde\rho^2) \\
2 & \frac{2}{3}(3+16\tilde\rho^2+16\tilde\rho^4)\\
3 & \frac{2}{15}(15+158\tilde\rho^2+384\tilde\rho^4+256\tilde\rho^6)\\
4 & \frac{2}{105}(105+1856\tilde\rho^2+8000\tilde\rho^4+12288\tilde\rho^6+6144\tilde\rho^8)\\
5 & \frac{2}{315}(315+8438\tilde\rho^2+56192\tilde\rho^4+146688\tilde\rho^6+163840\tilde\rho^8+65536\tilde\rho^{10})
\\ \hline
\mbox{} & p_E^m  \\
\hline
0 &  0  \\
1 & -4(1+\tilde\rho^2)   \\
2 & -\frac{16}{3}(1+3\tilde\rho^2+2\tilde\rho^4) \\
3 & -\frac{2}{15}(46+302\tilde\rho^2+512\tilde\rho^4+256\tilde\rho^6) \\
4 & -\frac{64}{105}(11+129\tilde\rho^2+406\tilde\rho^4+480\tilde\rho^6+192\tilde\rho^8) \\
5 & -\frac{4}{315}(563+10419\tilde\rho^2+52480\tilde\rho^4+108160\tilde\rho^6+98304\tilde\rho^8+32768\tilde\rho^{10})
\end{array}
\end{displaymath}
\caption{\label{Table:pKpE} The Polynomials $p_K^m$ and $p_E^m$ appearing in
\protect{Eq.\ (\ref{PhiPm})}. These are given in terms of the dimensionless
quantity $\tilde\rho\equiv\rho/(2P_{\varphi\varphi}^{1/2})$.}
\end{table}
\begin{table}
\centering
\begin{displaymath}
\begin{array}{c|l}
m & p_{1K}^m \\
\hline
0 & 0 \\
1 & -1 \\
2 & -8 \tilde\rho ^2-4\\
3 & \frac{1}{3} \left(-128 \tilde\rho ^4-128 \tilde\rho ^2-27\right)\\
4 & -\frac{16}{5} \left(64 \tilde\rho ^6+96 \tilde\rho ^4+42 \tilde\rho ^2+5\right)\\
5 & \frac{1}{35} \left(-32768 \tilde\rho ^8-65536 \tilde\rho ^6-44160 \tilde\rho
   ^4-11392 \tilde\rho ^2-875\right)
\\ \hline
\mbox{} & p_{1E}^m  \\
\hline
0 &  \frac{1}{2}  \\
1 & \tilde\rho ^2+\frac{1}{2}   \\
2 & 8 \tilde\rho ^4+8 \tilde\rho ^2+\frac{1}{2} \\
3 & \frac{1}{6} \left(256 \tilde\rho ^6+384 \tilde\rho ^4+134 \tilde\rho ^2+3\right)\\
4 &\frac{1}{10} \left(2048 \tilde\rho ^8+4096 \tilde\rho ^6+2496 \tilde\rho ^4+448
   \tilde\rho ^2+5\right)\\
5 & \frac{1}{70} \left(65536 \tilde\rho ^{10}+163840 \tilde\rho ^8+141568 \tilde\rho
   ^6+48512 \tilde\rho ^4+5318 \tilde\rho ^2+35\right)
\\ \hline\hline
\mbox{} & p_{2K}^m \\
\hline
0 & -1 \\
1 &-4 \tilde\rho ^2-2 \\
2 & \frac{1}{3} \left(-64 \tilde\rho ^4-64 \tilde\rho ^2-15\right)\\
3 &-\frac{2}{5} \left(256 \tilde\rho ^6+384 \tilde\rho ^4+178 \tilde\rho ^2+25\right)\\
4 & \frac{1}{105} \left(-49152 \tilde\rho ^8-98304 \tilde\rho ^6-68480 \tilde\rho
   ^4-19328 \tilde\rho ^2-1785\right)\\
5 & -\frac{2}{315} \left(327680 \tilde\rho ^{10}+819200 \tilde\rho ^8+765696 \tilde\rho
   ^6+329344 \tilde\rho ^4+63358 \tilde\rho ^2+4095\right)
\\ \hline
\mbox{} & p_{2E}^m  \\
\hline
0 & \tilde\rho ^2+\frac{1}{2}  \\
1 & 4 \tilde\rho ^4+4 \tilde\rho ^2+\frac{1}{2} \\
2 & \frac{1}{6} \left(128 \tilde\rho ^6+192 \tilde\rho ^4+70 \tilde\rho ^2+3\right)\\
3 & \frac{1}{10} \left(1024 \tilde\rho ^8+2048 \tilde\rho ^6+1288 \tilde\rho ^4+264
   \tilde\rho ^2+5\right) \\
4 & \frac{1}{210} \left(98304 \tilde\rho ^{10}+245760 \tilde\rho ^8+216832 \tilde\rho
   ^6+79488 \tilde\rho ^4+10322 \tilde\rho ^2+105\right)\\
5 & \frac{1}{630} \left(1310720 \tilde\rho ^{12}+3932160 \tilde\rho ^{10}+4455424
   \tilde\rho ^8+2357248 \tilde\rho ^6+574008 \tilde\rho ^4+50744 \tilde\rho
   ^2+315\right)
\\ \hline\hline
\mbox{} & p_{3K}^m \\
\hline
0 & -\frac{1}{24} \\
1 & \frac{1}{24} \left(-2 \tilde\rho ^2-1\right) \\
2 & \frac{1}{24} \left(16 \tilde\rho ^4+16 \tilde\rho ^2-1\right)\\
3 & \frac{32 \tilde\rho ^6}{3}+16 \tilde\rho ^4+\frac{21 \tilde\rho ^2}{4}-\frac{1}{24}\\
4 & \frac{1}{24} \left(2048 \tilde\rho ^8+4096 \tilde\rho ^6+2496 \tilde\rho ^4+448
   \tilde\rho ^2-1\right)\\
5 &\frac{1}{120} \left(65536 \tilde\rho ^{10}+163840 \tilde\rho ^8+142592 \tilde\rho
   ^6+50048 \tilde\rho ^4+5750 \tilde\rho ^2-5\right)
\\ \hline
\mbox{} & p_{3E}^m  \\
\hline
0 & \frac{1}{12} \left(2 \tilde\rho ^2+1\right)  \\
1 & \frac{1}{12} \left(\tilde\rho ^4+\tilde\rho ^2+1\right) \\
2 & \frac{1}{12} \left(-8 \tilde\rho ^6-12 \tilde\rho ^4-2 \tilde\rho ^2+1\right) \\
3 & \frac{1}{12} \left(-128 \tilde\rho ^8-256 \tilde\rho ^6-135 \tilde\rho ^4-7 \tilde\rho
   ^2+1\right)\\
4 &\frac{1}{12} \left(-1024 \tilde\rho ^{10}-2560 \tilde\rho ^8-2080 \tilde\rho ^6-560
   \tilde\rho ^4-14 \tilde\rho ^2+1\right)\\
5 & \frac{1}{60} \left(-32768 \tilde\rho ^{12}-98304 \tilde\rho ^{10}-106112 \tilde\rho
   ^8-48384 \tilde\rho ^6-7923 \tilde\rho ^4-115 \tilde\rho ^2+5\right)
\\ \hline\hline
\mbox{} & p_{4K}^m \\
\hline
0 & -\frac{1}{3} \\
1 & -\frac{4}{3} \left(2 \tilde\rho ^2+1\right) \\
2 & \frac{1}{3} \left(-64 \tilde\rho ^4-64 \tilde\rho ^2-13\right)\\
3 & -\frac{4}{15} \left(512 \tilde\rho ^6+768 \tilde\rho ^4+326 \tilde\rho ^2+35\right)\\
4 &\frac{1}{21} \left(-16384 \tilde\rho ^8-32768 \tilde\rho ^6-21632 \tilde\rho
   ^4-5248 \tilde\rho ^2-343\right)\\
5 &-\frac{4}{315} \left(327680 \tilde\rho ^{10}+819200 \tilde\rho ^8+738816 \tilde\rho
   ^6+289024 \tilde\rho ^4+45718 \tilde\rho ^2+1995\right)
\\ \hline
\mbox{} & p_{4E}^m  \\
\hline
0 & \frac{1}{6} \left(2 \tilde\rho ^2+1\right)\\
1 & \frac{1}{6} \left(16 \tilde\rho ^4+16 \tilde\rho ^2+1\right)\\
2 & \frac{64 \tilde\rho ^6}{3}+32 \tilde\rho ^4+11 \tilde\rho ^2+\frac{1}{6}\\
3 & \frac{1}{30} \left(4096 \tilde\rho ^8+8192 \tilde\rho ^6+4912 \tilde\rho ^4+816
   \tilde\rho ^2+5\right)\\
4 &\frac{1}{42} \left(32768 \tilde\rho ^{10}+81920 \tilde\rho ^8+69888 \tilde\rho
   ^6+22912 \tilde\rho ^4+2190 \tilde\rho ^2+7\right)\\
5 & \frac{1}{630} \left(2621440 \tilde\rho ^{12}+7864320 \tilde\rho ^{10}+8695808
   \tilde\rho ^8+4284416 \tilde\rho ^6+885936 \tilde\rho ^4+54448 \tilde\rho
   ^2+105\right)
\\ \hline
\end{array}
\end{displaymath}
\caption{\label{Table:pKpEn} The polynomials $p_{nK}^m$ and $p_{nE}^m$ appearing in
\protect{Eq.\ (\ref{Im})}. Here $\tilde\rho\equiv\rho/(2P_{\varphi\varphi}^{1/2})$.}
\end{table}

\section{The local source coefficients $\beta_0^m$} \label{AppB}

The coefficients $\beta_0^m$ in Eq.\ (\ref{beta}) are given by
\begin{equation}\label{beta0}
\beta_0^{m}=\frac{\hat\beta^m_0}{(r_0-2M)^2 \sqrt{P_{\varphi\varphi}}},
\end{equation}
where $\hat\beta^m_0$ are dimensionless polynomials in $M/r_0$. A list of these
polynomials, for $m=0$--$5$, is provided in Table \ref{Table:beta0}.
\begin{table}
\centering
\begin{displaymath}
\begin{array}{c|l}
m & \hat\beta_0^m \\
\hline\hline
0 &  -5/2   +12\zeta    -19\zeta^2  +8\zeta^3  \\
1 &  -7/2   +17\zeta    -25\zeta^2  +8\zeta^3  \\
2 &  -23/6  +56/3\zeta  -27\zeta^2  +8\zeta^3  \\
3 &  -121/30    +59/3\zeta  -141/5\zeta^2   +8\zeta^3  \\
4 &  -877/210   +428/21\zeta    -1017/35\zeta^2 +8\zeta^3  \\
5 &  -2701/630  +1319/63\zeta   -3122/105\zeta^2+8\zeta^3
\end{array}
\end{displaymath}
\caption{\label{Table:beta0} The dimensionless polynomials $\hat\beta^m_0$
appearing in \protect{Eq.\ (\ref{beta0})}. In this table $\zeta\equiv M/r_0$.}
\end{table}


\end{document}